%% file: main.tex
\documentclass[11pt]{article}
\input{header/packages}

\input{header/custom_defs}
\input{paper/title}

\input{header/new_defs}

\def\finalTreeBound{\poly(\Delta)}

\begin{document}

\maketitle

\input{paper/abstract}

\newpage 
\input{paper/introduction}

\input{paper/paper_main}

\end{document}

%% file: header/packages.tex
\usepackage[utf8]{inputenc}

\usepackage{fullpage}
\usepackage[margin = 2.5cm]{geometry}

\usepackage{amsmath, amssymb, amsthm}

\usepackage{graphicx}
\usepackage{color}
\usepackage[dvipsnames]{xcolor}

\usepackage{natbib}

\usepackage{nicefrac}

\usepackage{framed}
\usepackage[framemethod=tikz]{mdframed}
\usepackage{todonotes}
\usepackage{color}

\definecolor{darkgreen}{rgb}{0,0.5,0}
\definecolor{darkgray}{rgb}{0.2,0.2,0.2}
\usepackage{hyperref}
\hypersetup{
    unicode=false,          
    colorlinks=true,        
    linkcolor=blue,          
    citecolor=purple,        
    filecolor=magenta,      
    urlcolor=cyan           
}

\usepackage{algorithm}
\usepackage[noend]{algpseudocode}
\newcommand*{\longDefiningEquals}{\stackrel{\text{def}}{=\joinrel=}}
  
\usepackage{lineno}
\usepackage{caption}
\usepackage{enumerate}

\usepackage{tikzsymbols}
\usepackage{nicefrac}

\usepackage{subcaption}
\usepackage{pdfpages}
\usepackage{bbm}

\usepackage[section]{placeins}
\usepackage{soul}

\mdfdefinestyle{challenge}{
    linewidth=1pt,
    leftmargin=0,
    rightmargin=0,
    backgroundcolor=gray!5,
    innertopmargin=0pt,
    innerbottommargin=5pt,
}
\surroundwithmdframed[style=challenge]{challenge}

\usepackage{thmtools}

\usepackage{tcolorbox}



%% file: header/custom_defs.tex
\newtheorem{theorem}{Theorem}[section]
\newtheorem{claim}[theorem]{Claim}
\newtheorem{corollary}[theorem]{Corollary}
\newtheorem{lemma}[theorem]{Lemma}
\newtheorem{definition}[theorem]{Definition}

\newcommand{\bbZ}{\mathbb{Z}}

\newcommand{\calE}{\mathcal{E}}

\newcommand{\calI}{\mathcal{I}}

\newcommand{\calS}{\mathcal{S}}
\newcommand{\calT}{\mathcal{T}}

\newcommand{\calX}{\mathcal{X}}

\def\poly{\text{poly}}

\def\MIS{\mathsf{MIS}}
\def\eps{\varepsilon}

\usepackage{soul}

\def\MIS{\mathsf{MIS}}
\def\RG{\mathsf{RGMIS}}
\def\YES{\mathsf{YES}}
\def\NO{\mathsf{NO}}

\newtheorem{observation}[theorem]{Observation}

\newtheorem*{remark*}{Remark}


%% file: paper/title.tex
\title{Randomized Greedy MIS Is Fast at Every Node of a Tree}
\author{
Mina Dalirrooyfard \\ Morgan Stanley Research \and 
Konstantin Makarychev \\ Northwestern University \and 
Slobodan Mitrovi\'c \thanks{Supported by the Google Research Scholar and NSF Faculty Early Career Development Program No.~2340048. e-mail: \texttt{smitrovic@ucdavis.edu}} \\ UC Davis \and
Srikkanth Ramachandran \thanks{Supported by NSF Faculty Early Career Development Program No.~2340048. e-mail: \texttt{sramach@ucdavis.edu}} \\ UC Davis
}
\date{}

%% file: header/new_defs.tex
\usepackage{algorithm}
\usepackage[noend]{algpseudocode}
\usepackage{lineno}
\usepackage{caption}
\usepackage{framed}
\usepackage{enumerate}
\usepackage[capitalize, nameinlink]{cleveref}
\tcbuselibrary{skins,breakable}

\usepackage{tikzsymbols}
\usepackage{thmtools,thm-restate}
\usepackage{nicefrac}

\usepackage{subcaption}
\usepackage{pdfpages}
\usepackage{bbm}
\def\poly{\text{poly}}

\usepackage{mdframed}

\newtcolorbox[
  auto counter,
  number within=section,
  crefname={Alg}{Algs},
  Crefname={Alg}{Algs}
]{algobox}[2][]{%
  enhanced,
  breakable,
  title={Procedure~\thetcbcounter: #2},
  colback=gray!3,
  colframe=black,
  boxrule=0.8pt,
  arc=2mm,
  #1
}

\crefname{algobox}{Alg}{Algs}
\Crefname{algobox}{Alg}{Algs}

\crefname{goal}{Goal}{Goals}
\Crefname{goal}{Goal}{Goals}

\newtcolorbox[use counter=goal]{goalbox}[2][]{
    enhanced,
    breakable,
    colback=blue!3,
    colframe=blue!45!black,
    colbacktitle=blue!12,
    coltitle=black,
    title={Goal~\thetcbcounter: #2},
    boxrule=0.6pt,
    arc=2mm,
    left=3mm,
    right=3mm,
    top=2mm,
    bottom=2mm,
    label type=goal,
    #1
}

\usepackage{soul}
\crefname{theorem}{Theorem}{Theorems}
\crefname{problem}{Open Problem}{Open Problems}
\Crefname{lemma}{Lemma}{Lemmas}
\Crefname{invariant}{Invariant}{Invariants}
\Crefname{claim}{Claim}{Claims}
\Crefname{observation}{Observation}{Observations}
\Crefname{algorithm}{Algorithm}{Algorithms}
\Crefname{figure}{Figure}{Figures}

\usepackage{enumitem}

\newlist{defitems}{enumerate}{3}

\setlist[defitems,1]{
  label=(\arabic*),
  ref=\thedefinition(\arabic*),
  leftmargin=*,
  itemsep=0.2em
}
\crefname{defitemsi}{Def}{Defs}
\Crefname{defitemsi}{Def}{Defs}

\setlist[defitems,2]{
  label=(\roman*),
  ref=\thedefinition(\arabic{defitemsi}.\roman*),
  leftmargin=*,
  itemsep=0.2em
}

\setlist[defitems,3]{
  label=(\alph*),
  ref=\thedefinition(\arabic{defitemsi}.\roman{defitemsii}.\alph*),
  leftmargin=*,
  itemsep=0.2em
}

\crefname{defitemsi}{Def}{Defs}
\Crefname{defitemsi}{Def}{Defs}

\crefname{defitemsii}{Def}{Defs}
\Crefname{defitemsii}{Def}{Defs}

\crefname{defitemsiii}{Def}{Defs}
\Crefname{defitemsiii}{Def}{Defs}

\newlist{thmitems}{enumerate}{1}
\setlist[thmitems]{label=(\alph*),ref=\thetheorem(\alph*)}
\crefname{thmitemsi}{Theorem}{Theorems}
\Crefname{thmitemsi}{Theorem}{Theorems}

\newlist{obsitems}{enumerate}{1}
\setlist[obsitems]{label=(\alph*),ref=\theobservation(\alph*)}
\crefname{obsitemsi}{Obs}{Obs}
\Crefname{obsitemsi}{Obs}{Obs}

\newlist{lemitems}{enumerate}{1}
\setlist[lemitems]{label=(\alph*),ref=\thelemma(\alph*)}
\crefname{lemitemsi}{Lemma}{Lemmas}
\Crefname{lemitemsi}{Lemma}{Lemmas}

\newlist{claimitems}{enumerate}{1}
\setlist[claimitems]{label=(\alph*),ref=\theclaim(\alph*)}
\crefname{claimitemsi}{Claim}{Claims}
\Crefname{claimitemsi}{Claim}{Claims}

\newlist{algsteps}{enumerate}{1}
\setlist[algsteps]{label=\arabic*.,ref=\thetcbcounter Step.(\arabic*)}
\crefname{algstepsi}{Alg}{Alg}
\Crefname{algstepsi}{Alg}{Alg}

%% file: paper/abstract.tex
\begin{abstract}

The Randomized Greedy MIS (RGMIS) algorithm computes a maximal independent set (MIS) by processing the vertices in a uniformly random order and greedily adding each vertex to the independent set whenever possible. 
Despite its simplicity, this algorithm has proved effective in a range of parallel, distributed, and dynamic settings.
For sufficiently small maximum degree, Nguyen and Onak (FOCS~'08) and Yoshida, Yamamoto, and Ito (STOC~'09) showed that a recursive simulation of RGMIS can be used to estimate the sizes of matching and vertex cover in sublinear time.
The key subroutine in their approach is the following local query procedure: given a vertex $v$, to determine whether $v$ belongs to the MIS, recursively evaluate the neighbors of $v$ that precede $v$ in the random order.
If any such neighbor is found to belong to the MIS, the procedure immediately concludes that $v$ is not in the MIS; otherwise, it concludes that $v$ is in the MIS.
For graphs of maximum degree $\Delta$, Nguyen and Onak showed that, in expectation over the random ordering, this procedure makes $O(\exp(\Delta))$ recursive calls from a worst-case starting vertex.
No better upper bound is known even for trees.
We prove that on trees, from \textit{any} starting vertex, this recursive RGMIS procedure makes $\mathrm{poly}(\Delta)$ recursive calls in expectation.
\end{abstract}

\newpage

%% file: paper/introduction.tex
\section{Introduction}
\vspace{-5pt}
In a graph $G=(V,E)$, a set of vertices $I \subseteq V$ is \emph{independent} if no two vertices in $I$ are adjacent. A \emph{maximum} independent set is an independent set of largest possible cardinality, while a \emph{maximal} independent set is one that is inclusion-wise maximal. 
Finding a maximum independent set is computationally intractable in general~\cite{karp1972reducibility}, whereas a maximal independent set can be computed by a simple greedy algorithm in linear time. 
Throughout the paper, MIS stands for \emph{maximal} independent set.

In 2008, Nguyen and Onak~\cite{nguyen2008constant} initiated the study of determining, in sublinear time, whether a queried vertex $u \in V$ belongs to the MIS produced by the following random-order greedy process:\footnote{The approach was used by \cite{nguyen2008constant} in the context of computing a maximal matching.}
\begin{itemize}
	\item For each $u \in V$, assign rank $r(u)$ chosen uniformly at random from $[0, 1]$;
	\item Greedily augment independent set $I$ by iterating over $V$ from smallest to largest vertex with respect to their ranks.
\end{itemize}
We refer to this random-order greedy process as randomized greedy MIS (RGMIS).
Nguyen and Onak then studied the following recursive simulation of RGMIS: to determine whether $u$ belongs to the MIS, one recursively evaluates every neighbor $w$ of $u$ with $r(w) < r(u)$. 
Then $u$ belongs to the MIS if and only if none of these neighbors do.
Letting $\Delta$ denote the maximum degree of the input graph, the authors proved that the expected number of recursive calls made by this procedure is $e^{O(\Delta)}$.
In fact, this upper bound is tight even on trees.\footnote{A fixed path of length $L$ appears in a stack of recursive calls with probability $1/L!$. So, the expected number of recursive calls from the root of a $\Delta$-ary tree on $n$ vertices is $\sum_{L = 0}^{\log_\Delta n} \Delta^L / L! = \Theta(e^\Delta)$ for a sufficiently large $n$.}
The authors also observed that a natural search-pruning heuristic may asymptotically improve this recursive process: if, for some neighbor $w$ of $u$ with $r(w) < r(u)$, the recursive call returns that $w$ belongs to the MIS, then it immediately follows that $u$ does not belong to the MIS, and hence no other neighbor of $u$ needs to be explored.
We refer to RGMIS implemented with this heuristic as pruned RGMIS (see Algorithm \ref{alg:rgmis}).
The authors conjectured that pruned RGMIS is significantly faster than the unpruned one, and left that direction for future research.

A year later, Yoshida, Yamamoto, and Ito~\cite{yoshida2009improved} proved that this pruning step drastically improves the \emph{average-case} complexity of RGMIS.
More precisely, they showed that when the query vertex is chosen uniformly at random, pruned RGMIS makes only $|E(G)|/|V(G)|$ recursive calls in expectation.
In particular, on trees, the expected number of recursive calls made by pruned RGMIS from a uniformly random query vertex is less than $1$.
Very recently, Dalirrooyfard, Makarychev, and Mitrović~\cite{dalirrooyfard2026simple} gave a considerably simpler analysis of the result of~\cite{yoshida2009improved}.

However, works~\cite{yoshida2009improved,dalirrooyfard2026simple} leave open whether pruned RGMIS may still require $\exp(\Theta(\Delta))$ recursive calls from a worst-case vertex. This raises the question of whether, at least on trees, the average-case and worst-case complexities of pruned RGMIS coincide. Kapralov, Mitrović, Norouzi-Fard, and Tardos~\cite{kapralov2020space} ruled this out by showing that the two can differ substantially.
More formally, they show that for every constant $\eps > 0$, there exist a tree and a vertex $u$ in that tree such that pruned RGMIS invoked on $u$ makes $\Omega_{\eps}(\Delta^{2-\eps})$ recursive calls in expectation.\footnote{Their lower-bound is for computing a maximal matching on trees, which is equivalent to computing MIS on the line-graph of an input tree. Nevertheless, in the lower-bound construction, the authors optimize the computation; see Section~7.1 of the arXiv version of \cite{kapralov2020space}. As a byproduct, their lower bound applies to RGMIS on trees as well.}
Thus, even on trees, pruned RGMIS exhibits a substantial gap between average-case and worst-case behavior: the average complexity is constant, while the worst-case complexity is at least polynomial in $\Delta$. This already shows that the tree case captures a nontrivial aspect of the problem.


Although Ghaffari~\cite{ghaffari2022local} gives a local computation algorithm for MIS with query complexity $\poly(\Delta)$ in expectation\footnote{The same work also gives a $\poly(\Delta)\log n$ query LCA that is correct with high probability.}, this does not address the complexity of the specific recursive process induced by RGMIS. Studying this process is natural in its own right: RGMIS is an especially simple and canonical randomized greedy procedure, and related versions of it have been studied in several areas beyond local computation. 
For instance, random-order greedy processes of this kind long predate their local-computation applications: closely related versions were studied already in Flory's 1939 work in chemistry~\cite{flory1939intramolecular}, and later in the random sequential adsorption literature~\cite{evans1993random}.
It has also served as a useful primitive in a number of later algorithmic works.

Since~\cite{nguyen2008constant}, techniques for analyzing RGMIS and related randomized greedy processes have found applications in local and sublinear-time algorithms for vertex cover, matching, set cover, and minimum dominating set~\cite{nguyen2008constant,yoshida2009improved,onak2012near,behnezhad2022time}. Related ideas have also appeared in subsequent work on dynamic and sublinear-time matching, as well as in several other algorithmic settings~\cite{bhattacharya2023dynamic,behnezhad2023dynamic,mahabadi20250,mahabadi2025sublinear,mahabadi2026sublinear,varma2023average,behnezhad2022almost,azarmehr2025stochastic,das2026history}. However, these developments do not seem to yield progress on the expected query complexity of RGMIS from a worst-case queried vertex. A closely analogous difficulty was noted explicitly by Behnezhad~\cite{behnezhad2022time} in the context of randomized greedy maximal matching, where they wrote that ``\emph{analyzing the expected query-complexity for a \emph{given} edge seems to be challenging.}'' In the RGMIS setting, the analogous question concerns a worst-case queried vertex.

The question we study goes back to Nguyen and Onak~\cite{nguyen2008constant}, who conjectured in their concluding remarks that their exponential analysis might be far from tight, writing that the complexity may grow ``quasi-polynomially, not exponentially, in $\Delta$.'' Despite subsequent progress, this question remains open even on trees.
\begin{center}
    \emph{What is the expected complexity of pruned RGMIS from a worst-case vertex in a tree?}
\end{center}

\subsection{Our result}
In this paper we focus on the pruned RGMIS, and hence refer to it as RGMIS for simplicity.
We say that a vertex $x$ \emph{queries} a neighbor $y$ if the recursive execution of RGMIS at $x$ invokes RGMIS on $y$.
Given an input tree and a query vertex $v$, we root the tree at $v$. Thus, every recursive call proceeds from a vertex to one of its children, and every \emph{query path} is simply a path of successive queries starting at $v$.

The run-time of RGMIS is within $O(\Delta)$ factor of the size of the recursion tree produced by the RGMIS algorithm (see Algorithm \ref{alg:rgmis}). The $\Delta$ factor comes from each recursive call having to inspect the adjacency list of a new vertex in the tree and initiating subsequent recursive calls. The run-time bound is also a bound on the portion of the input tree that is inspected by the algorithm. Thus, up to polynomial factor, it suffices to bound the size of the recursion tree, which we henceforth call the \textit{query tree}, produced by the RGMIS algorithm.

We prove that the expected size of the $\RG$ query tree over a worst-case starting vertex (see Definition \ref{defn:query:tree}) obtained from the RGMIS algorithm is polynomial in $\Delta$ on trees.

\begin{restatable}{theorem}{maintheorem}
\label{main:thm}
    There exists an absolute constant $C > 0$ such that for every integer $\Delta \geq 2$, every tree $\calT$ with maximum degree $\Delta$ and every vertex $u$ in $\calT$, the expected size of the query tree $Q(\calT, u, r)$ from $u$ over uniformly assigned random ranks $r$ is at most $\Delta^C$, i.e.,
    \[        \mathbbm{E}_{r}\bigl[|Q(\calT, u, r)|\bigr] \leq \Delta^C.
    \]
\end{restatable}
\subsection{Other related work}
A closely related body of work arises in the framework of local computation algorithms (LCAs), where one seeks to answer local queries while remaining consistent with a single global solution. The LCA model was introduced by Rubinfeld, Tamir, Vardi, and Xie~\cite{rubinfeld2011fast} and further developed by Alon, Rubinfeld, Vardi, and Xie~\cite{alon2012space}. 
Maximal independent set has been one of the central benchmark problems in this area. 
In particular, beyond the recursive simulation of randomized greedy orders considered in the present line of work, subsequent papers obtained improved LCAs for MIS and related problems, including deterministic and stateless algorithms~\cite{even2018best}, algorithms for non-constant-degree graphs~\cite{levi2015local}, and, very recently, an LCA for MIS with query complexity $\poly(\Delta)\log n$~\cite{ghaffari2022local}. 
More broadly, many other problems have been studied in the centralized local model as well; we refer the reader to the survey of Levi and Medina~\cite{levi2017centralized} for a general overview. 

There is also a separate probabilistic line of work that studies the random greedy independent set via branching-process and local-limit methods, especially on random trees and sparse random graphs. 
This includes analyses on regular graphs of large girth~\cite{gamarnik2010randomized}, random graphs with prescribed degree sequences~\cite{brightwell2017greedy}, and more recent frameworks that explicitly involve Galton--Watson trees or related branching objects~\cite{krivelevich2024greedy,contat2022surprising}. 
These works are close in spirit, since they also analyze the random greedy MIS process, but their focus is typically on the size or density of the resulting independent set in random graph models, or on the behavior seen from a typical root, rather than on the worst-case query complexity from a fixed vertex in an arbitrary tree. 
For this reason, it is not clear that their techniques directly apply to the setting studied here.

\subsection{Overview of our approach}

As already discussed, without pruning, running unpruned RGMIS from the root of a $\Delta$-ary tree makes $\Theta(e^\Delta)$ queries in expectation.
Hence, to show that pruned RGMIS has significantly lower complexity, the main task is to quantify the effect of pruning. To illustrate our starting idea, consider the following thought experiment.

Suppose we want to count the number of recursive calls made at vertices with ranks in $(1/2,1)$, and ignore all work performed at vertices of rank below $1/2$. Moreover, we count a recursive call from $u$ to $v$ only if $v$ itself queries some vertex $w$, and the ranks of $u$, $v$, and $w$ all lie in $(1/2,1)$. How many vertices with ranks in $(1/2,1)$ are then queried?

Consider a vertex $u$ with $r(u) \in (1/2,1)$ that is reached by the recursion. At that point, among all randomness in the subtree rooted at $u$, only the rank $r(u)$ has been revealed. Suppose that $u$ queries a child $v$, which in turn queries some vertex $w$ with $r(w) \in (1/2,1)$. For this to happen, it must be the case that all children $w'$ of $v$ with $r(w') < 1/2$ are not in the MIS. Otherwise, the RGMIS procedure would stop exploring the children of $v$ upon reaching such a vertex $w'$, and as a result would never query $w$.

We thus consider the set of \emph{promising} vertices $v$, defined as follows: let $A$ be the set of children $v$ of $u$ such that no child $w$ of $v$ with $r(w) < 1/2$ belongs to the MIS. \emph{In expectation, how many vertices in $A$ with ranks at least $1/2$ are queried by $u$?}
To answer this, observe that membership in $A$ is determined by the subtree below each $v \in A$, and therefore does not reveal any information about $r(v)$ itself.
Hence, conditioned on $A$, each rank $r(v)$ for $v \in A$, remains uniformly distributed in $[0,1]$.
A naive estimate therefore gives only that the expected number of vertices in $A$ with rank at least $1/2$ queried by $u$ is at most $|A|/2$.
\emph{However, pruning gives a much stronger bound.}
Indeed, suppose that some $v \in A$ satisfies $r(v) \le 1/2$. Then $r(v) < r(u)$, so $u$ queries $v$ before any child $v'$ with $r(v') > 1/2$. Moreover, by the definition of $A$, no child of $v$ with smaller rank belongs to the MIS, and hence $v$ itself belongs to the MIS. It follows that $u \notin \MIS$, so the exploration stops at $u$ before any child of $u$ with rank larger than $1/2$ is queried.
Therefore, children in $A$ with rank at least $1/2$ are queried only if every vertex in $A$ has rank at least $1/2$. Since the ranks $\{r(v) : v \in A\}$ remain independent and uniform in $[0,1]$, this event occurs with probability $2^{-|A|}$.
Thus, conditioned on $A$, the expected number of queried vertices in $A$ with ranks in $(1/2,1)$ is at most $2^{-|A|}\cdot |A|$ which is at most a constant.

This thought experiment captures the basic mechanism behind our proof, but turning it into a full proof requires overcoming two obstacles.
First, what about the cost of exploring vertices with ranks below $1/2$, which we have so far ignored?
Second, and a major challenge in our work, how can one continue this reasoning recursively once previously revealed randomness creates dependencies that were absent in the definition of $A$?
We address these obstacles in two stages.

\paragraph{Interval decomposition, encodings, and the first obstacle (Section~\ref{sec:first-section}).}
We begin with the first obstacle, namely, that the thought experiment ignores all work performed at vertices of rank below $1/2$.
To handle this, we partition the rank space into intervals of the form $[2^{-j},2^{-j+1})$ and use these intervals to encode paths according to the scale of the ranks they encounter.
For a fixed such encoding, we then enlarge the class of paths we consider and work with \emph{candidate paths}, namely paths whose vertices lie in the prescribed intervals, even if they are not necessarily queried by the recursion.
Thus, the interval decomposition is what both organizes the analysis and defines the larger family of paths on which we work.

The point of passing to candidate paths is that they reveal only coarse information about the ranks along the path -- namely, the interval in which each rank lies, but not its exact value inside that interval.
This is exactly the amount of information we can afford to reveal.
It preserves enough randomness so that the relative order of vertices that fall within the same interval remains hidden. As a result, the probability that a candidate path is a query path is roughly $1/(k_1!\cdots k_m!)$, where $k_i$ denotes the number of vertices on the candidate path that lie in the interval $(1/2^i,\,1/2^{i-1})$.

\paragraph{Refined encodings and the second obstacle (Section~\ref{sec:candidate-paths}).}
The second obstacle is that the ``clean'' independence used in the thought experiment breaks down under recursion.
In the thought experiment, the set $A$ was useful because we could determine membership in $A$ without revealing the actual ranks of the vertices in $A$ themselves.
In the recursive process, however, once we move one level deeper, some information in the relevant subtrees has already been revealed.
Thus, even after passing to candidate paths, one cannot simply repeat the same argument at every level.

This is where the main technical tool of the proof enters, namely a \emph{refined path encoding}.
At each step, there is a natural threshold coming from the interval of the next vertex on the path.
The refined encoding records how much this threshold needs to be ``lowered'', and uses \emph{flags} to mark a sparse set of levels where such lowering is genuinely necessary.
Informally, these are the levels where we must pay for a large branching event; at the other levels, we avoid revealing comparable information.
This is what allows the thought-experiment intuition to be applied repeatedly without exposing too much randomness along the way.
Combined with the exponential tail bound for such large-branching events -- which we already saw informally in the thought experiment -- this yields an exponentially decaying contribution for each valid encoding.
Summing over all encodings then gives the desired bound on the number of candidate paths.

\subsection{Statement on AI use}

All of the text in this paper was written by the (human) authors of this submission. AI was used only for proofreading: to identify grammatical and mathematical errors. All comments and suggestions made by AI were thoroughly reviewed by the authors. All proofs and technical ideas in this paper were developed solely by the authors, without any AI involvement.

%% file: paper/paper_main.tex
\section{Preliminaries}

The randomized greedy MIS algorithm is outlined in Algorithm \ref{alg:rgmis}.

\begin{algorithm}
    \begin{algorithmic}[1]
        \Procedure{$\RG$}{$G, u, r$}
        \Statex \textbf{Input:} A graph $G$, vertex $u$ and distinct ranks $r : V(G) \rightarrow [0, 1)$.
        \Statex \textbf{Output:} Does $u$ belong to the $\MIS$ in the greedy algorithm that processes the vertices in increasing order of $r$?
            \State Obtain $\Gamma(u)$, the list of neighbors of $u$. \label{line:qry:adjList}
            \For{each $w \in \Gamma(u)$ with $r(w) < r(u)$ in increasing order of $r(w)$}
                \If{$\RG(G, w, r) = \YES$}
                    \State \Return $\NO$ \label{line:prune}
                \EndIf 
            \EndFor 
            \State \Return $\YES$
        \EndProcedure
    \end{algorithmic}
    \caption{The $\RG$ algorithm}
    \label{alg:rgmis}
\end{algorithm}

In this paper, we are interested in the (expected) size and depth of the recursion tree produced by Algorithm \ref{alg:rgmis} starting from a \textit{worst-case} vertex $v$ of a tree, where expectation is taken over uniformly sampled vertex ranks.

\begin{definition}[Query tree and size]
\label{defn:query:tree}
    Given a graph $G$, vertex $u \in V(G)$ and rank $r : V(G) \rightarrow [0, 1)$, we denote by $Q(G, u, r)$, the recursion tree obtained from the execution of $\RG(G, u, r)$.

    We use $|Q(G, u, r)|$ to denote the size (number of vertices) of the tree $Q(G, u, r)$. 
\end{definition}

\begin{definition}[Query path]
\label{defn:query:path}
    We say that a path $P = (u_0, u_1, \dots u_{L-1})$ in a tree $\calT$ is a query path if the execution of $\RG(\calT, u_i, r)$ directly invokes $\RG(\calT, u_{i+1}, r)$ for each $i = 0,1,\dots L-2$. For simplicity, we say that $u_i$ queries $u_{i+1}$ if $\RG(\calT, u_i, r)$ invokes $\RG(\calT, u_{i+1}, r)$. 
\end{definition}

\begin{definition}[Worst case expected query complexity]
\label{defn:worst-query}
    For an integer $\Delta \geq 1$, let $\mathrm{Tree}(\Delta)$ denote the set of all trees of maximum degree $\Delta$. 
    The worst case expected query complexity of Algorithm \ref{alg:rgmis} on a tree of maximum degree $\Delta$ is given by,
    \[ 
        \max\limits_{\substack{G \in \mathrm{Tree}(\Delta), \\ v \in G}} \mathbbm{E}_{r} \Big[ |Q(G, v, r)| \Big]
    \]
\end{definition}

\paragraph{Technicality of distinct ranks.} In Definition \ref{defn:worst-query}, we assume that the ranks $r : V(G) \rightarrow [0, 1)$ are sampled uniformly and independently for each vertex $v \in V(G)$. However, Algorithm \ref{alg:rgmis} requires that the ranks are distinct for each vertex. In our analysis we will assume that $r$ is distinct for all vertices, which is an almost sure event. Hence, our analysis also applies to the version of $\RG$ that samples a uniform permutation $\pi$ and sets the rank as $r(u) = \pi(u) / n$ where $\pi(u) \in \{0, 1, \dots n-1\}$.

\section{Counting the number of query paths}
\label{sec:first-section}
In this section, we give the proof of Theorem \ref{main:thm} assuming a technical lemma (Lemma \ref{lem:bound-candidate-paths}). We prove Lemma \ref{lem:bound-candidate-paths} in Section \ref{sec:candidate-paths}. 
We start by introducing some definitions and explaining the intuition behind them.

\paragraph{Setup and notation} 
Consider a rooted tree $\calT$ and let $u_0$ be its root. Consider the execution of $\RG(\calT, u_0, r)$ over uniformly sampled rank $r : V(\calT) \rightarrow [0, 1)$. We denote by $\calT_v$, the sub-tree beneath vertex $v$ in the rooted tree. We denote by $\Gamma(v)$ the set of children of $v$ in the rooted tree $\calT$. More formally, $\Gamma(v)$ is the set of neighbors of $v$ that are farther than $v$ from $u_0$.

We show that the expected number of query paths starting at the root is at most $\finalTreeBound$, where $\Delta$ is the maximum number of children of any vertex in the tree when rooted at $u_0$. Observe that the maximum degree is at most $\Delta + 1$ and hence the $\poly(\Delta)$ claim also holds with $\Delta$ being the maximum degree in the tree.
Our strategy will be to partition the interval $[0, 1)$ into sub-intervals with geometrically decreasing sizes.

For an integer $j$, let $\tau(j) = 2^{-j}$. We partition the interval $[0,1)$ into infinitely many intervals $[\tau(j),\tau({j-1}))$, for $j \in \bbZ_{> 0}$. Every number $x \in (0,1)$, and in particular (w.p. 1), the rank of every vertex, belongs to exactly one such interval. Thus, for every query path $P = (u_0, u_1, \dots u_{L-1})$, there exists a unique choice of numbers $s_0,\dots,s_{L-1} \in \bbZ_{> 0}$ such that,
\begin{equation}\label{eq:agrees-with-s}
r(u_i) \in [\tau({s_i}),\tau({s_i-1})) \quad\quad\quad \text{for all } i \in \{0, 1, \dots L-1\}.
\end{equation}
Here $r(u)$ denotes the rank of vertex $u$. If a path $P = (u_0,\dots,u_{L-1})$ satisfies~\eqref{eq:agrees-with-s}, then we say that $P$ agrees with $s$. Note that for every \emph{query} path $P$ we have
$r(u_0) > r(u_1) > \dots > r(u_{L-1})$, and therefore the corresponding sequence satisfies $s_0 \le s_1 \le \dots \le s_{L-1}$.

Let $m = \lceil \log_2 \Delta \rceil$. The suffix of the path with $s_i > m$ have ranks smaller than $1/\Delta$. For such vertices, pruning has negligible effect and the number of monotone paths is in itself small enough. In Section \ref{sec:low-rank-paths}, we show that the contribution by such vertices can only increase the expected value by a constant factor. Hence, in this main technical portion of the paper, we count only those paths ending at vertices with rank at least $\tau(m)$. Thus, we are mainly interested in paths that agree with non-decreasing sequences $s$ containing integers in $[m]$. We use $[k]$ to denote the set $\{1, 2, \dots k\}$. The relevant sequences of length $L$ are:
\[
\calS(L) = \{ s \in [m]^L : s_0 \le s_1 \le \dots \le s_{L-1} \}.
\]

We denote by $\calS$ the union of $\calS(L)$ over all $L \geq 0$. To upper bound the expected number of query paths, we separately bound the number of query paths that agree with each $s \in \calS(L)$ and then sum these bounds over all $s \in \calS(L)$ and all $L \ge 1$.

We now introduce some notation. We denote the depth of a vertex $u$ in the tree by $\delta(u)$, where the depth of the root is $0$. Note that if $u_0,\dots, u_{L-1}$ is a query path starting at the root, then $\delta(u_i) = i$. The interpretation of~\eqref{eq:agrees-with-s} is that the rank of a vertex $u$ on a downward path from the root that agrees with $s$ must lie in the interval $[\tau({s_{\delta(u)}}),\tau({s_{\delta(u)}-1}))$. We will denote the maximal independent set (MIS) obtained by the RGMIS algorithm by $\calI$. This set is random and depends on the choice of ranks of vertices in the tree.
We now give two key definitions. These definitions form the crux of our proof and will enable us to expose and reveal required randomness to prove our claims.

\subsection{Definition of Promising Vertices}
The first definition is the notion of a \textit{promising vertex}. The reader should carefully note the specific way in which it is defined and the randomness that it depends on.
\begin{definition}[Promising vertices]
\label{def:promising-set}
We say that a vertex $v$ is \emph{promising} at threshold $\theta$
if no child $w \in \Gamma(v)$ with $r(w) < \theta$ satisfies,
\[
    \RG(\calT_w, w ,r) = \YES
\]
That is, in the execution of $\RG$ in the sub-tree $\calT_w$ beneath $w$ using the same rank function $r$, the vertex $w$ is added to the independent set.
\end{definition}

\def\prom{\text{Prom}}
The event $\prom(v, \theta) \longDefiningEquals$ ``$v$ is promising at threshold $\theta$'' is random and is determined by the random ranks of the vertices in the tree. Note however that $\prom(v, \theta)$ is intentionally defined to be \textit{independent} of $r(v)$ itself, but rather depends only on the ranks in the strict sub-forest $\calT_v - v$. The next crucial observation is that although $\prom(v, \theta)$ inspects only ranks in $\calT_v - v$, all the vertices $w \in \calT_v - v$ that are actually queried by the RGMIS algorithm have the \textit{same output} in the sub-tree $\calT_w$ as the global tree $\calT$, which we prove below in Observation \ref{obs:local-global-exec-equiv}. In particular, for our purposes, we may assume\footnote{In the cases that the conclusion doesn't hold, vertex $v$ does not lie in the query tree of interest.} that $\prom(v, \theta)$ implies that $v$ does not contain a child $w$ with $r(w) < \theta$ and $w \in \calI$. We prove this in Lemma \ref{lem:promis-motiv}. 

\begin{observation}[Local-Global execution equivalence]
\label{obs:local-global-exec-equiv}
    Suppose $\RG(\calT, u_0, r)$ invokes, either directly or indirectly, $\RG(\calT, v, r)$ for some vertex $v \in \calT$. It holds that,
    \[
        \RG(\calT, v, r) = \RG(\calT_v, v, r)
    \]
    That is, the output of the execution of vertex $v$ in the global tree $\calT$ is equivalent to the output of $v$ in the sub-tree $\calT_v$ beneath vertex $v$ in $\calT$.
\end{observation}
\begin{proof}
    We prove the stronger statement that the RGMIS recursion trees generated by the two executions are identical. It follows from the rule of RGMIS, i.e., output at $v$ is the NOR of the outputs of all neighbors of $v$ in the query tree (with NOR of an empty set being $\YES$), that the outputs are identical.

    First if $v = u$, then the statement is trivial. Henceforth we may assume that $v \neq u$ and hence $p_v$, the parent of $v$ in the rooted tree $\calT$ exists and is unique.
    Since $\RG(\calT, v, r)$ is invoked, it holds that $r(p_v) > r(v)$. Hence $\RG(\calT, v, r)$ does not invoke $\RG(\calT, w, r)$ for any $w \in \calT - V(\calT_w)$. A consequence of this is that all the recursive calls generated directly or indirectly by $\RG(\calT, v, r)$ are of the form $\RG(\calT, w, r)$ for some $w \in \calT_v$. All of these vertices are also present in $\calT_v$ and hence this is a valid RGMIS query tree for the call $\RG(\calT_v, v, r)$ which is unique and hence must also be identical to the query tree for that call. It follows that,
    \[
        \RG(\calT, v, r) = \RG(\calT_v, v, r)
    \]
    as desired.
\end{proof}

\begin{lemma}[Promising definition motivation]
\label{lem:promis-motiv}
    Consider an arbitrary vertex $v$ in the tree that is queried by $\RG(\calT, u_0, r)$. If $v$ is promising at threshold $\theta$ and $r(v) < \theta$, then $v$ belongs to the global independent set $\calI$ (w.p. $1$).
\end{lemma}
\begin{proof}
    Consider all the vertices $w$ with $r(w) < r(v)$ that is queried by the recursive call $\RG(\calT, v, r)$. From Observation \ref{obs:local-global-exec-equiv}, it holds that
    \[
        \RG(\calT, w, r) = \RG(\calT_w, w, r)
    \]
    As $r(w) < r(v) < \theta$, the definition of promising gives, $\RG(\calT_w, w, r) = \NO$.
    Combining with the equation above, we get $\RG(\calT, w, r) = \NO$. From the pseudo-code of the RGMIS algorithm (See Algorithm \ref{alg:rgmis}), it follows that $\RG(\calT, v, r) = \YES$, i.e., $v \in \calI$.
\end{proof}

\subsection{Definition of Candidate Paths and Candidate Sets}

The second key ingredient will be the definition of \textit{Candidate Paths} and closely related notion of \textit{Candidate Sets}. Given a path $P = (u_0, u_1, \dots u_{L-1})$ and a sequence $s$, we say that $P$ is a candidate path for $s$ if for each valid $i$, (i) $r(u_i)$ belongs to $[\tau({s_i}), \tau(s_i - 1))$ and more crucially, (ii) $u_i$ is promising at threshold $\tau(s_{i+1})$. The event (ii) captures the effect of pruning in the algorithm. It guarantees that in the local tree $\calT_{u_i}$, the vertex $u_i$ has not terminated execution by finding an MIS vertex with rank smaller than $\tau(s_{i+1})$. By the intuition presented in the previous sub-section, this implies that if $u_i$ is queried by the algorithm, then the recursive call $\RG(\calT, u_i, r)$ cannot have terminated execution before querying a vertex with rank smaller than $\tau(s_{i+1}) \leq r(u_{i+1})$, which is a conservative condition to be satisfied if $u_i$ is to query $u_{i+1}$ after implementing pruning. 
We next formally state this definition.

\begin{definition}[Candidate Path]
\label{def:cand-path}
A downward path $u_0,\dots,u_{L-1}$ starting at the root is a candidate path agreeing with $s$ if, (i) it agrees with $s$ and, (ii) for all $i < L-1$, $u_{i}$ is promising at threshold $\tau(s_{i+1})$.
\end{definition}

From the observations in the previous sub-section it is not hard to show that a query path $P$ with ranks agreeing with $s$ must also be a candidate path for $s$. We shall defer the proof for now (See Lemmas \ref{lem:candidate-path},\ref{cl:every-v-promising}). We next define candidate sets for vertices $u_i$ that contain all possible choices for $u_{i+1}$ assuming that the prefix $(u_0, u_1, u_2, \dots u_i)$ is a potential prefix of a candidate path for $s$.

\begin{definition}[Candidate Set]
\label{def:cand-set}
Let $s \in \calS(L)$ and $\theta \in (0,1)$. For every vertex $u$, we define the candidate set $A_{\theta}(u)$ as follows. If $u$ is not promising at $\theta$, then we set $A_{\theta}(u) = \varnothing$. Otherwise,
\[
A_{\theta}(u) = \{ v \in \Gamma(u) : r(v) \in [\tau(s_{\delta(v)}),\tau(s_{\delta(v)}-1)) \text{ and } v \text{ is promising at threshold } \theta \}.
\]
\end{definition}

\begin{observation}[Candidate paths can be encoded by candidate sets]
    If $(u_0, u_1, \dots u_{L-1})$ is a candidate path for sequence $s$, then for each $i < L-2$ and $\theta \leq \tau(s_{i+2})$, 
    \[u_{i+1} \in A_{\theta}(u_i)\] 
    That is, the prefix $(u_0, u_1, \dots u_{L-2})$ of the path may be encoded by an integer sequence $(a_1, a_2, \dots a_{L-2})$ where $a_{i+1}$ is the position of $u_{i+1}$ in the set $A_{\theta}(u_i)$ for any $\theta \leq \tau(s_{i+2})$
\end{observation}
\begin{proof}
    Since $u_i$ is promising at threshold $\tau(s_{i+1})$, $A_{\theta}(u_i)$ it is not forcibly set to $\varnothing$ by Definition \ref{def:cand-set} for any $\theta \leq \tau(s_{i+1})$ which includes all $\theta \leq \tau(s_{i+2})$. Now, $u_{i+1} \in A_{\theta}(u_i)$ for $\theta \leq \tau(s_{i+2})$ now holds from (i) $P$ agreeing with $s$ and (ii) $u_{i+1}$ being promising at threshold $\tau(s_{i+2})$ and hence also $\theta$.
\end{proof}

\begin{observation}[Membership in $A_{\theta}(u)$ is monotone in $\theta$]
\label{obs:cand-set-mono-theta}
    For any vertex $u$ and any $\theta, \theta'$ with $\theta' \leq \theta$ we have,
    \[A_{\theta}(u) \subseteq A_{\theta'}(u)\]
\end{observation}

\begin{proof}
If a vertex $v$ is promising at threshold $\theta \in (0,1)$, then it is also promising at any threshold $\theta' < \theta$.
First, we consider the case $A_{\theta'}(u) = \emptyset$ due to $u$ not being promising at $\theta'$. It follows that $u$ is not promising at $\theta$ and hence $A_{\theta}(u) = A_{\theta'}(u)=\varnothing$. Henceforth, we assume $u$ is promising at $\theta'$.

Next, for an arbitrary $v \in A_{\theta}(u)$, suppose $v$ is promising at $\theta$. It holds that $v$ is also promising at $\theta'$. Hence $v \in A_{\theta'}(u)$. We have $A_{\theta}(u) \subseteq A_{\theta'}(u)$ as desired.
\end{proof}

We now show that, for a query path agreeing with $s$, each next vertex $v$ belongs to the candidate set of the preceding vertex $u$ at threshold $\tau({s_{\delta(v)}})$, with the exception of the last two vertices. We begin by proving that every vertex $v$ on such a query path, except possibly the last one, is promising at threshold $\tau({s_{\delta(v)}})$.

\begin{claim}\label{cl:every-v-promising}
Consider a path $u_0,\dots,u_{L-1}$ and $s \in \calS(L)$. If this path is a query path agreeing with~$s$, then each $u_i$ with $i < L-1$ is promising at threshold $\tau({s_{{i+1}}})$. 
\end{claim}

\begin{proof}
Suppose $P = u_0,\dots,u_{L-1}$ is a query path agreeing with $s$, and fix $i < L-1$.
Since $P$ is a query path, vertex $u_i$ queries $u_{i+1}$. Because $P$ agrees with $s$,
the rank of $u_{i+1}$ lies in $[\tau({s_{i+1}}),\tau({s_{i+1}-1}))$ and, in particular,
$\tau({s_{i+1}}) \le r(u_{i+1})$.

Suppose, for contradiction, that there exists a child $w \in \Gamma(u_i)$ such that
$r(w) < \tau({s_{i+1}})$ and $\RG(\calT_w, w, r) = \YES$. Since $r(w) < r(u_{i+1})$ and $u_i$ queries $u_{i+1}$, from the pseudocode of RGMIS it follows that $w$ is called by $u_i$. Now, Observation \ref{obs:local-global-exec-equiv} applies to $w$, giving $\RG(\calT, w, r) = \YES$, i.e., $w \in \calI$. Since RGMIS queries children
in increasing order of rank and stops once it encounters a child in $\calI$, vertex
$u_i$ would stop querying its children upon reaching $w$. As $r(w) < \tau({s_{i+1}}) \le r(u_{i+1})$,
this would prevent $u_i$ from querying $u_{i+1}$, contradicting the assumption that
$P$ is a query path.
Therefore, no such vertex $w$ exists, and $u_i$ is promising at threshold $\tau({s_{i+1}})$.
\end{proof}

\begin{lemma}\label{lem:candidate-path}
If $P = (u_0,\dots,u_{L-1})$ is a query path agreeing with $s$, then for every $i < L-2$ and $\theta \le \tau({s_{i+2}})$
it holds that,
\[
u_{i+1} \in A_{\theta}(u_i).
\]
\end{lemma}

\begin{proof}
Fix an index $i < L-2$ and consider the vertex $u_i$. We first show that the first case in the definition of $A_{\theta}(u_i)$ does not occur. Suppose, for contradiction, that $u_i$ is not promising at threshold $\theta$. As $u_i$ is queried, this implies that $u_i$ contains a child $w$ with $r(w) < \theta$ and $\RG(\calT_w, w, r) = \YES$. As $r(w) < r(u_{i+1})$, $w$ must be queried. Since $w$ is queried, then by Observation \ref{obs:local-global-exec-equiv}, $\RG(\calT, w, r) = \YES$. 
From the pseudocode of RGMIS, $u_i$ must terminate before it can query any vertex with rank at least $\theta$ (either by terminating at $w$ or an even earlier vertex) which includes $u_{i+1}$ as $r(u_{i+1}) \geq \tau(s_{i+1}) \geq \tau(s_{i+2}) \geq \theta$. This contradicts the assumption that $P$ is a query path.

It remains to show that (i) $r(u_{i+1}) \in [\tau(s_{i+1}), \tau(s_{i+1}-1))$ and (ii) $u_{i+1}$ is promising at threshold $\theta$. Condition (i) is guaranteed by the assumption that $P$ agrees with $s$. To prove condition (ii) we show that $u_{i+1}$ is promising at threshold $\tau(s_{i+2}) \geq \theta$ which is stronger.

Consider an arbitrary child $w \in \Gamma(u_{i+1})$ with $r(w) < \tau(s_{i+2}) \leq r(u_{i+2})$.
Since $P$ is a query path, $u_{i+1}$ queries $u_{i+2}$. From the psuedocode of RGMIS this implies that $u_{i+1}$ also queries $w$ as it has higher priority during the recursive calls.
Using Observation \ref{obs:local-global-exec-equiv}, since $w$ is queried by the algorithm, we have $\RG(\calT, w, r) = \RG(\calT_{w}, w, r)$. Since $u_{i+1}$ does not terminate at $w$, $\RG(\calT, w, r) = \NO$. It follows that $\RG(\calT_w, w, r) = \NO$ and that $v$ is promising at threshold $\tau(s_{i+2})$ and hence also $\theta \leq \tau(s_{i+2})$.
\end{proof}

The above lemmas imply that every query path agreeing with $s$ is also a candidate path agreeing with~$s$. 
Importantly, the converse does not hold, and the conditional probability that a path is a query path given that it is a candidate path is relatively small. The following lemma provides an upper bound on this probability.

\begin{lemma}[Conditional probability of query paths given candidate path]
\label{lem:prob-candidate-path-is-query}
Fix $s \in \calS(L)$. Let $k_1,\dots,k_{m}$ denote the number of indices $i$ such that $s_i = 1,\dots,m$, respectively. Then, for every fixed path $P$, we have
\[
\Pr\bigl(P \text{ is a query path} \mid 
P \text{ is a candidate path agreeing with } s\bigr)
\le \frac{1}{k_1!\,k_2!\dots k_{m}!}.
\]
\end{lemma}

We now prove the lemma. The key step is to justify the following observation. The proof of Lemma \ref{lem:prob-candidate-path-is-query}  then follows as the ranks of vertices in the query paths must be montone. 

\begin{lemma}\label{lem:uniform-indep}
Fix a downward path $P = u_0,\dots,u_{L-1}$ and a sequence $s \in \calS(L)$. Conditioned on the event that $P$ is a candidate path agreeing with $s$, the random variables $r(u_0),\dots,r(u_{L-1})$ are mutually independent, and for each $i$ the rank $r(u_i)$ is uniformly distributed over $[\tau({s_i}),\tau({s_i-1}))$.
\end{lemma}

\begin{proof}
It is sufficient to show that if $P$ is a candidate path agreeing with $s$
under ranks~$r$, then $P$ is also a candidate path for any set of ranks~$r'$
such that  
(i) $r'(u_i) \in [\tau({s_i}),\tau({s_i-1}))$ for all $i$, and  
(ii) $r'(v) = r(v)$ for all $v \notin P$.

To establish this, it suffices to prove that under $r'$, for each $i < L-1$, $u_{i+1}$ is promising at threshold $\tau(s_{i+1})$. The other condition is that $r'(u_i) \in J_{s_i}$ which holds by construction.
Consider the two
sets
\begin{align*}
W &= \{\, w \in \Gamma(u_i) : r(w) < \tau({s_{i+1}})\,\},\\
W' &= \{\, w \in \Gamma(u_i) : r'(w) < \tau({s_{i+1}})\,\}.
\end{align*}

We first argue that $W = W'$. If $w \in W$, then $w \notin P$, since its depth is
$i+1$, while the only vertex on $P$ at depth $i+1$ is $u_{i+1}$, which satisfies
$r(u_{i+1}) \in [\tau({s_{i+1}}),\tau({s_{i+1}-1}))$, whereas $r(w) < \tau({s_{i+1}})$. Hence
assumption~(ii) implies $r'(w) = r(w)$, so $w \in W'$. The converse direction is
identical, giving $W = W'$.

Next observe that since every $w \in W$ lies outside the path $P$, every vertex
in the subtree rooted at $w$ also lies outside $P$. Therefore, by assumption~(ii),
all ranks in this subtree are identical under $r$ and $r'$, and hence the behavior
of RGMIS within this subtree is the same under $r$ and $r'$. It follows that $\RG(\calT_w, w, r) = \NO$ for each $w \in W$ which suffices to prove that $u_{i}$ is promising at threshold $\tau(s_{i+1})$ under rank $r'$ as desired.
\end{proof}

We have shown that, for any fixed sequence $s \in \calS(L)$, the probability that a given downward path $P$ is a query path agreeing with $s$, conditioned on the event that it is a candidate path agreeing with $s$, is small. Specifically, Lemma~\ref{lem:prob-candidate-path-is-query} bounds this conditional probability by a factor depending only on the multiplicities of the values appearing in $s$. Thus, in order to bound the expected number of query paths, it suffices to control the expected number of candidate paths.

The main technical step of the analysis is therefore to bound the expected number of candidate paths of length $L$ agreeing with a fixed sequence $s$. In the next subsection, we prove the following key result.

\begin{lemma}\label{lem:bound-candidate-paths}
There exists absolute constants $C_1 > 0$, $C_2 > 0$ such that for every $s \in \calS(L)$, the expected number of candidate paths of length $L$ agreeing with $s$ is at most
$
f(L, \Delta) = C_1^L \cdot \Delta^{C_2}
$
.
\end{lemma}

Given Lemma \ref{lem:bound-candidate-paths} Theorem \ref{main:thm} follows by an application of linearity of expectation over query paths agreeing with all sequences in $\calS$. We restate the theorem here for convenience and give its proof assuming Lemma \ref{lem:bound-candidate-paths}.

\maintheorem*

\begin{proof}[Proof of Theorem \ref{main:thm} given Lemma \ref{lem:bound-candidate-paths}]
\label{proof-of-reduction}
Fix $L \ge 1$ and a sequence $s \in \calS(L)$, and let $k_1,\dots,k_{m}$ denote the number of indices $i$ such that $s_i = 1, \dots, m$, respectively. By Lemma~\ref{lem:prob-candidate-path-is-query}, the expected number of query paths of length $L$ agreeing with $s$ is at most
\[
f(L,\Delta)\cdot \frac{1}{k_1!\cdots k_{m}!}.
\]
Summing over all $L\ge 1$ and all admissible multiplicity vectors $(k_1,\dots,k_{m})$ with $k_1+\cdots+k_{m}=L$, we obtain the following upper bound on the expected number of query paths:
\[
\sum_{L \ge 1}\sum_{\substack{k\in \bbZ_{\geq 0}^m:\\ k_1+\cdots+k_{m}=L}}
f(L,\Delta)\cdot \frac{1}{k_1!\cdots k_{m}!}
=\sum_{L \ge 1}
f(L,\Delta)\cdot 
\sum_{\substack{k\in \bbZ_{\geq 0}^m:\\ k_1+\cdots+k_{m}=L}} \frac{1}{k_1!\cdots k_{m}!}.
\]
We now show that the inner sum on the right-hand side equals $m^L/L!$.

\begin{claim}\label{cl:sum-inv-factorials}
For any pair of integers $m \ge 1$ and $L \ge 0$,
\[
\sum_{\substack{k\in \bbZ_{\geq 0}^m:\\ k_1+\cdots+k_{m}=L}} \frac{1}{k_1!\cdots k_{m}!}
=\frac{m^L}{L!}.
\]
\end{claim}
\begin{proof}
Consider the multivariate generating function
\[
G(x_1,\dots,x_{m}) = e^{x_1}\cdots e^{x_{m}}.
\]
On the one hand, writing each exponential as its Taylor expansion and taking the product, we obtain
\[
G(x_1,\dots,x_{m})
= \prod_{j=1}^{m}\Bigl(\sum_{k_j \ge 0} \frac{x_j^{k_j}}{k_j!}\Bigr)
= \sum_{k_1,\dots,k_{m} \ge 0}
\frac{x_1^{k_1}\cdots x_{m}^{k_{m}}}{k_1!\cdots k_{m}!}.
\]
Therefore, the sum of the coefficients of all monomials of total degree $L$ in this expansion is exactly
\[
\sum_{\substack{k\in \bbZ_{\geq 0}^m:\\ k_1+\cdots+k_{m}=L}}
\frac{1}{k_1!\cdots k_{m}!}.
\]
On the other hand, since $e^{x_1}\cdots e^{x_{m}} = e^{x_1+\cdots+x_{m}}$, substituting $x_1=\cdots=x_{m}=x$ yields
\[
G(x,\dots,x) = e^{mx}.
\]
The coefficient of $x^L$ in the Taylor expansion of $e^{mx}$ is $m^L/L!$. This coefficient is exactly the sum of the coefficients of all monomials of total degree $L$ in the multivariate expansion of $G$, since each such monomial becomes $x^L$ under the substitution $x_1=\cdots=x_{m}=x$.
Combining the two expressions for this coefficient completes the proof.
\end{proof}

Using Claim~\ref{cl:sum-inv-factorials}, we bound the expected number of query paths agreeing with some $s \in \calS$ by, 
\begin{align*}
\sum_{L \ge 1} f(L,\Delta)\cdot \frac{m^L}{L!} &= 
\Delta^{C_2} \cdot \sum_{L \ge 1} (C_1)^L\cdot \frac{m^L}{L!}
=
\Delta^{C_2} \cdot \sum_{L \ge 1} \frac{(C_1 \cdot m)^L}{L!}  \\&=
\Delta^{C_2} \cdot (e^{C_1 \cdot m} - 1) < (\Delta)^{(\log_2 e) \cdot C_1 + C_2}.
\end{align*}
and obtain the desired upper bound on the expected number of query paths agreeing with sequences $s \in \calS$. It remains to argue about paths to vertices with rank smaller than $\tau(m)$. For such vertices, the probability that the rank falls in the range $[0, \tau(m))$ is at most $\Delta^{-1}$, so by linearity of expectation, the expected number of such monotone paths from every vertex is at most $\sum\limits_{L \geq 0} \Delta^L \cdot \Delta^{-L} / L! = e$. One can consider these paths to be independent of whether the root of the path is queried by $\RG$ as Algorithm \ref{alg:rgmis} is essentially a depth-first search on the tree and when a vertex $v$ is visited for the first time, none of its descendants have been visited and their ranks uninspected.
In Section \ref{sec:low-rank-paths}, Lemma \ref{lem:deal-with-low-rank} we show formally that such paths can only increase the expected value by a factor of $e$. Finally, observe that for every vertex $v$ in the query tree, the unique path from $u_0$ to $v$ is a query path. Conversely for every query path $P$, the end of the path $P$ is a vertex in the query tree. Thus the number of query paths is equivalent to the size of the query tree.
\end{proof}

\section{Upper bound on the number of candidate paths}
\label{sec:candidate-paths}
In this section we prove Lemma \ref{lem:bound-candidate-paths} which is the most technical part of the paper. Recall that Lemma \ref{lem:bound-candidate-paths} bounds the expected number of candidate paths agreeing with a fixed $s \in \calS(L)$. At a high level, the argument is as follows. We show that the probability that the size of the candidate set $A_\theta(u)$ exceeds a constant is small. Thus, for a typical vertex $u$, the set $A_\theta(u)$ contains only a constant number of children of $u$. If we had $|A_\theta(u)| \le O(1)$ for all vertices $u$, then we could upper bound the number of candidate paths by $\exp{(O(L))}$, which would conclude the proof.

Unfortunately, for some vertices $u$ the size $|A_\theta(u)|$ may be large, and moreover the sizes of $A_\theta(u')$ and $A_\theta(u'')$ are not necessarily independent for $u' \neq u''$. In this subsection, we show how to nevertheless upper bound the expected number of candidate paths agreeing with $s$.

We begin by upper bounding the probability that $|A_\theta(u)|$ is large. The bound holds even if the ranks of some children of $u$ are
partially revealed.

\begin{lemma}\label{lem:prob-ub-large-candidate-set}
Let $s \in \calS(L)$. For every vertex $u$ in the tree with $\delta(u) < L-1$, every $\theta \le \tau({s_{\delta(u)+2}})/2$, and every $T > 1$, the following holds. We define $\calT_{\theta}(u)$ to be the set of the following random variables.
\[
    \calT_{\theta}(u) = \{r(v) \mid v \in \calT_w \text{ for some } w \in \Gamma(u) \text{ with } r(w) < \theta\}
\]
Let $\cal E$ be any event that depends only on the ranks in $\calT_{\theta}(u)$. Note that this implies that every child $w \in \Gamma(u) \setminus \calT_{\theta}(u)$ has $r(w) \geq \theta$ and that no other information about the ranks of vertices in $\calT_w$ is needed to determine $\cal E$.  Then it holds that,
\begin{equation}
\label{eq:lem:prob-ub-large-candidate-set}
\Pr\bigl(|A_{2\theta}(u)| \geq T \mid {\cal E}\bigr)
\le \exp\!\left(-\,\frac{T}{\frac{\tau({s_{\delta(u)+1}})}{\theta} + 1}\right).
\end{equation}
\end{lemma}
In Lemma~\ref{lem:prob-ub-large-candidate-set}, $T$ is an arbitrary positive number. However, we may assume without loss of generality that $T$ is an integer by replacing it with $\lceil T \rceil$. 
Indeed, this does not change the probability on the left-hand side, since $|A_{2\theta}(u)|$ is always integral, while the expression on the right-hand side only decreases. 
Hence, from now on we assume that $T \in \bbZ_{\ge 0}$.

To prove this lemma, we first detour through a small probabilistic puzzle. Imagine a game in which a player repeatedly draws (with replacement) from a bag containing three types of tokens labeled $1$, $2$, and $3$. Each draw is independent: the probability of drawing a $1$ is $p_1$, of drawing a $2$ is $p_2$, and of drawing a $3$ is $p_3$, with $p_1 + p_2 + p_3 = 1$. The player makes $n$ draws.

We are interested in the probability that, among these $n$ draws, the player observes at least $T$ tokens labeled $1$ and no tokens labeled $2$. This probability clearly depends on $T$, $n$, and the parameters $p_1, p_2, p_3$. However, we would like a clean upper bound that depends only on $T$ and on the ratio $\lambda = p_1/(p_1 + p_2)$. Thus, the question is: for fixed $T$ and $\lambda$, and arbitrary $p_1, p_2, p_3$ satisfying $\lambda = p_1/(p_1 + p_2)$ and $n \ge T$, how large can this probability be?

It is not hard to see (and we will prove formally) that the probability is maximized when there are no $3$'s and we take exactly $n = T$ draws; that is, when $p_1 = \lambda$, $p_2 = 1 - \lambda$, $p_3 = 0$, and $n = T$. In this case, the event ``at least $T$ ones and no twos'' simply means ``all $T$ draws are $1$'', which has probability $\lambda^T$.

In the remainder of the argument, we show that $\lambda^T$ is indeed an upper bound for every admissible choice of $p_1, p_2, p_3$ and every $n \ge T$. There are several ways to derive this inequality---for example, via a standard martingale or stopping-time argument---but here we present a more elementary probabilistic proof.

\begin{claim}\label{cl:multinomial-tail}
Let $X_1,\dots,X_n$ be i.i.d.\ random variables taking values in $\{1,2,3\}$ with
$\Pr(X=1)=p_1$, $\Pr(X=2)=p_2$, and $\Pr(X=3)=p_3$, where $p_1+p_2+p_3=1$.
Then for any positive integer $T$,
\[
\Pr\Bigl(\bigl|\{i : X_i=1\}\bigr| \ge T
\ \text{and}\
\bigl|\{i : X_i=2\}\bigr| = 0\Bigr)
\le
\Bigl(\frac{p_1}{p_1+p_2}\Bigr)^T .
\]
\end{claim}

\begin{proof}
Let $\calX = \{ i : X_i \in \{1,2\} \}$. Then the event
\[
\bigl|\{i : X_i=1\}\bigr| \ge T
\quad\text{and}\quad
\bigl|\{i : X_i=2\}\bigr| = 0
\]
is equivalent to the event that $|\calX| \ge T$ and $X_i = 1$ for all $i \in \calX$.

We condition on the random set $\calX$. Conditional on $\calX$, the random variables
$\{X_i : i \in \calX\}$ are i.i.d.\ with
\[
\Pr(X_i = 1 \mid i \in \calX) = \frac{p_1}{p_1+p_2}.
\]
Therefore, the conditional probability that $X_i = 1$ for all $i \in \calX$ and $|\calX| \ge T$ is
\[
\Bigl(\frac{p_1}{p_1+p_2}\Bigr)^{|\calX|}
\mathbf{1}\{|\calX| \ge T\}
\le
\Bigl(\frac{p_1}{p_1+p_2}\Bigr)^T.
\]
Taking expectation over $\calX$ completes the proof.
\end{proof}

\begin{proof}[Proof of Lemma~\ref{lem:prob-ub-large-candidate-set}]
Let $Y_u$ denote the set of all children of $u$ that are promising at threshold $2\theta$. This is a random set. Observe that $Y_u$ is independent of the ranks of the children of $u$, since the event that a vertex $v$ is promising at threshold $2\theta$ is completely determined by the ranks of vertices in the subtree rooted at $v$, excluding the rank of $v$ itself.

The candidate set $A_{2\theta}(u)$ is empty if there exists a vertex $v \in Y_u$ whose rank lies in the interval $[0,2\theta)$\footnote{Said vertex $v$ satisfies $\RG(\calT_v, v, r) = \YES$, and $u$ is not promising at $2\theta$ which nullifies $A_{2\theta}(u)$.}. If $A_{2\theta}(u)$ is non-empty then,

\[
A_{2\theta}(u) = \{ v \in Y_u : r(v) \in [\tau(s_{\delta(u)+1}),\tau(s_{\delta(u)+1}-1)) \}.
\]
Hence the event $|A_{2\theta}(u)| \geq T$ for $T > 0$ implies that there are at least $T$ vertices $v \in Y_u$ with
\[
r(v) \in [\tau(s_{\delta(u)+1}),\tau(s_{\delta(u)+1}-1)),
\]
and no vertices $v \in Y_u$ with
\[
r(v) \in [0,2\theta).
\]

We condition on the set $Y_u$. If there exists a vertex $v \in Y_u$ whose rank
is fixed by the event $\cal E$ and satisfies $r(v) < \theta$, then
$r(v) \in [0,2\theta)$, and hence $A_{2\theta}(u) = \varnothing$. Otherwise,
if no such vertex exists, then, conditioned on $\calT_\theta(u) \cup Y_u$, the ranks
$\{r(v) : v \in Y_u\}$ are i.i.d.\ uniformly distributed over $[\theta,1)$. Note that $\calE$ is determined by $\calT_\theta(u)$, so the above conditioning is stronger than the conditioning on $\calE$ and any bound we derive holds for the conditioning $\calE \cup Y_u$ as well.

We now apply Claim~\ref{cl:multinomial-tail}. For each $v \in Y_u$, define
\[
X_v =
\begin{cases}
1, & \text{if } r(v) \in [\tau({s_{\delta(u)+1}}),\tau({s_{\delta(u)+1}-1})),\\
2, & \text{if } r(v) \in [\theta, 2\theta),\\
3, & \text{otherwise}.
\end{cases}
\]
Conditional on the set $Y_u$, the random variables $\{X_v : v \in Y_u\}$ are i.i.d. in $[\theta, 1)$, with
\[
p_1 = \Pr(X_v = 1) = \tau({s_{\delta(u)+1}}) \cdot (1 - \theta)^{-1},
\qquad
p_2 = \Pr(X_v = 2) = \theta \cdot (1 - \theta)^{-1},
\]
and $p_3 = 1 - p_1 - p_2$.

By Claim~\ref{cl:multinomial-tail}, we have
\[
\Pr\Bigl(
\bigl|\{v \in Y_u : X_v = 1\}\bigr| > T
\ \text{and}\
\bigl|\{v \in Y_u : X_v = 2\}\bigr| = 0
\;\Big|\; Y_u
\Bigr)
\le
\Bigl(\frac{p_1}{p_1+p_2}\Bigr)^T.
\]
Since
\[
\frac{p_1}{p_1+p_2}
=
\frac{\tau({s_{\delta(u)+1}})}{\tau({s_{\delta(u)+1}})+\theta}
=
1-\frac{1}{\frac{\tau({s_{\delta(u)+1}})}{\theta}+1},
\]
it follows that
\[
\Bigl(\frac{p_1}{p_1+p_2}\Bigr)^T
=
\Bigl(1-\frac{1}{\frac{\tau({s_{\delta(u)+1}})}{\theta}+1}\Bigr)^T
\le
\exp\!\left(
-\frac{T}{\frac{\tau({s_{\delta(u)+1}})}{\theta}+1}
\right).
\]

This bound holds uniformly over all realizations of $Y_u$. Removing the conditioning on $Y_u$, we conclude that
\[
\Pr\bigl(|A_{2\theta(u)}| \geq T
\mid \calE\bigr) 
\le 
\Pr\bigl(|A_{2\theta(u)}| \geq T
\mid \calT_{\theta}(u)\bigr) 
\le
\exp\!\left(
-\frac{T}{\frac{\tau({s_{\delta(u)+1}})}{\theta}+1}
\right),
\]
as claimed.
\end{proof}

\subsection{Path Encoding}

We now define a path encoding. To this end, we first introduce some notation. We begin by fixing an arbitrary but consistent ordering of the children of each
vertex. Specifically, for every vertex $u$, we assume that the children in
$\Gamma(u)$ are numbered from $0$ to $|\Gamma(u)|-1$. While the particular
ordering is not important, we require it to be fixed once and for all. 

Next, for the fixed sequence $s$, we define the \emph{scale ratio} at level $i$ by
\[
\rho_s(i) = \frac{\tau({s_i})}{\tau(s_{i+1})} = 2^{s_{i+1}-s_i}.
\]
Since $s$ is non-decreasing, we have $s_i \le s_{i+1}$ for all $i$, and
consequently $\rho_s(i) \ge 1$.

We also define a sequence $\{\kappa_z\}_{z \ge 0}$ by
\[
\kappa_z = 5\cdot (z+1) \cdot 2^{z+1}.
\]

The encoding describes how to traverse a candidate path by specifying,
at each step, how the next vertex is selected from an appropriate candidate
set. For each step $i$, the encoding includes three parameters: $z_i$,
which determines the threshold used to define the candidate set; $a_i$,
which selects a particular child from this set; and $f_i$, which serves as
an auxiliary flag used to control steps where the candidate set is unusually large.

\paragraph{Intuition.}
In the encoding, we use a sequence of candidate sets corresponding to decreasing values of the threshold $\theta_i$. The candidate set indexed by $z$ is defined using the threshold $\theta_i = \tau(s_{i+2}+z)$. We use the candidate set with index $z$ only if its size is bounded by approximately $\kappa_z$ (more precisely, $\kappa_z \rho_s(i+1)$). In particular, if we use the candidate set with $z=0$, then the number of options for $a_i$ is bounded by $\kappa_0 = 10$ (up to the factor $\rho_s(i+1)$). This choice of $z$ is therefore preferable for the encoding.
However, it is not always feasible: if the candidate set at this threshold is too large, we must increase $z$ and decrease the threshold in order to satisfy the required size bound. Lowering the threshold enlarges the candidate set, and thus increases the number of possible values of $a_i$, but allows the construction to proceed.

To control such events, we later apply Lemma~\ref{lem:prob-ub-large-candidate-set} to argue that, in a $\kappa$-bounded encoding representing at least one candidate path, it is unlikely that the sum of the values $z_i$ is large. To facilitate this analysis, we use the flags $f_i$ to mark positions where the threshold must be decreased. More precisely, $f_i = 1$ indicates that the threshold $\theta_i$ cannot be increased (i.e., $z_i$ cannot be reduced) without violating the size bound on the candidate set. This corresponds to the condition
$|A_{2\theta_i}(u_i)| > \kappa_{z_i-1}\rho_s(i+1)$
(when $z_i \ge 1$).

\begin{definition}[$\kappa$-bounded path encoding]\label{def:path-encoding}
A \emph{$\kappa$-bounded path encoding} of length $L$ is a sequence of triplets
\[
(z_0, a_0, f_0), \dots, (z_{L-3}, a_{L-3}, f_{L-3}),
\]
where $z_i \in \bbZ_{\ge 0}$, $a_i \in \bbZ_{\ge 0}$, and $f_i \in \{0,1\}$.
For each $i < L-2$, define
\[
\theta_i = \tau(s_{i+2}+z_i).
\]
The triplets must satisfy the following conditions:
\begin{enumerate}
\item\label{cond:item:a} For each $i$, we have
\[
a_i \leq \kappa_{z_i}\,\rho_s(i+1).
\]
\item\label{cond:item:b} If $f_i = 1$, then $z_i > 0$.
\item\label{cond:item:c} If $f_i = 1$ and $i\geq 1$, then $\theta_{i-1} \leq \theta_{i}$ and $f_{i-1}=0$. 

\item\label{cond:item:d} For all $i \le L-3$, if $z_i > 0$, then at least one of the following holds:
\begin{enumerate}
\item\label{subcond:item:a} $f_i = 1$;
\item\label{subcond:item:b} $f_{i-1} = 1$ and $z_i \le z_{i-1}$; or
\item\label{subcond:item:c} $f_{i+1} = 1$ and $z_i \le z_{i+1} + s_{i+3} - s_{i+2}$.
\end{enumerate}
To simplify the treatment of boundary conditions, we set $f_{-1} = 0$ and $f_{L-2} = 0$.

\noindent\textbf{Remark.}
If $f_{i+1} = 1$, then by condition (C) we have $\theta_i \le \theta_{i+1}$, and hence
$s_{i+2} + z_i \ge s_{i+3} + z_{i+1}$. This implies that in item~D.c.\ the
inequality can be replaced by equality, i.e.,
$
z_i = z_{i+1} + s_{i+3} - s_{i+2}
$.
\end{enumerate}
\end{definition}

\begin{definition}\label{def:enc-represents}
We say that a $\kappa$-bounded path encoding $\{(z_i,a_i,f_i)\}$ \emph{represents}
the candidate path $u_0,\dots,u_{L-1}$ agreeing with a sequence $s$ if, for every $i < L-2$,
the following conditions hold:
\begin{enumerate}
\item\label{def:enc-represents:a} $u_{i+1}$ is the $a_i$-th element of $A_{\theta_i}(u_i)$, where the set is ordered according to the fixed ordering of the children, with indices starting from $1$;

\item\label{def:enc-represents:b} if $f_i = 1$, then
\[
|A_{2\theta_i}(u_i)| > \kappa_{z_i-1}\,\rho_s(i+1),
\]
where $\theta_i = \tau(s_{i+2}+z_i)$. 
\end{enumerate}
\end{definition}

We prove that every candidate path is represented by at least one
$\kappa$-bounded path encoding, and that each encoding represents at most $\Delta$
candidate paths. The $\Delta$ factor appears because the Decode\&Verify procedure decodes all vertices except for the last.
We then show that the probability that a fixed encoding
$\{(z_i,a_i,f_i)\}$ corresponds to a candidate path is at most
$\poly(\Delta) \cdot \exp(-\sum z_i)$.

\begin{lemma}
Every candidate path is represented by at least one
$\kappa$-bounded path encoding.
\end{lemma}
\begin{proof}
Consider a candidate path $u_0,\dots,u_{L-1}$. For every $i < L-2$,
let $z^{\star}_i$ be the smallest integer such that for every $z \ge z^{\star}_i$,
the following bound holds:
\begin{equation}\label{eq:def-z-star}
|A_{\tau({s_{i+2}+z)}}(u_i)| \le \kappa_{z}\rho_s(i+1).   
\end{equation}

Note that such $z^{\star}_i$ exists, since for sufficiently large $z$ we have
$\kappa_{z} \ge \Delta$, while $|A_{\tau({s_{i+2})\;+\;z}}(u_i)| \le \Delta$,
as $u_i$ has at most $\Delta$ children.

We construct a subset $F \subseteq \{0, 1, \dots L-2\}$ satisfying the following properties:
\begin{itemize}
\item if $i \in F$, then $z^{\star}_i > 0$ and $i+1 \notin F$;
\item if $z^{\star}_i > 0$, then at least one of the following holds:
(1) $i \in F$; 
(2) $i-1 \in F$ and $z^{\star}_{i-1} \ge z^{\star}_i$; or
(3) $i+1 \in F$ and $z^{\star}_{i+1} \ge z^{\star}_i$.
\end{itemize}

\begin{claim}
There exists a set $F$ satisfying the above properties.
\end{claim}

\begin{proof}
We construct $F$ greedily as follows. Consider all indices $i$ with
$z^{\star}_i > 0$, sorted in non-increasing order of $z^{\star}_i$. Process these indices one by one, and add $i$ to $F$ if neither of its
neighbors, $i-1$ nor $i+1$, has already been added to $F$. 

By construction, $F$ is an independent set, so if $i \in F$, then
$i+1 \notin F$, establishing the first property. For the second property,
fix $i$ with $z^{\star}_i > 0$. If $i \in F$, we are done. Otherwise,
$i$ was not added because at least one of its neighbors was already in $F$;
let $j \in \{i-1,i+1\}$ be such a neighbor. Since $j$ was processed no later
than $i$, we have $z^{\star}_j \ge z^{\star}_i$, establishing the claim.
\end{proof}

We now construct an encoding $(z_0,a_0,f_0),\dots,(z_{L-3},a_{L-3},f_{L-3})$ using a set $F$ as above.
For every $i \in F$, we set $f_i = 1$ and let $z_i = z^{\star}_i$. For $i \notin F$, we set $f_i = 0$. If $i \notin F$ and $i+1 \notin F$, we let $z_i = z^{\star}_i$; otherwise, if $i \notin F$ and $i+1 \in F$, we let
$z_i = \max(z^{\star}_i,z^{\star}_{i+1} + s_{i+3} - s_{i+2})$. As in the definition of a $\kappa$-bounded path encoding, we set $\theta_i = \tau(s_{i+2}+z_i)$. Finally, we let $a_i$ be the index of vertex $u_{i+1}$ in the candidate set $A_{\theta_i}(u_i)$ according to the fixed ordering of the children of $u_i$.

We now verify that the constructed encoding is valid and that it represents the candidate path $(u_0,\dots,u_{L-2})$. We begin by checking that the size constraints are satisfied. First observe that $z_i \ge z^{\star}_i$ for all $i$. Thus, by the definition of $z^{\star}_i$, we have
\[
|A_{\theta_i}(u_i)| \le
|A_{\tau({s_{i+2}+z_i})}(u_i)| \le \kappa_{z_i}\rho_s(i+1).
\]
Consequently,
\[
a_i \le \kappa_{z_i}\rho_s(i+1).
\]

Next, we verify the monotonicity condition on the thresholds. Suppose that $f_i = 1$ and $i \ge 1$. Then $f_{i-1} = 0$ by the properties of $F$. Moreover,
\[
z_{i-1} = \max\bigl(z^{\star}_{i-1},\, z^{\star}_i + s_{i+2} - s_{i+1}\bigr)
\ge z^{\star}_i + s_{i+2} - s_{i+1}
= z_i + s_{i+2} - s_{i+1}.
\]
It follows that $s_{i+1} + z_{i-1} \ge z_i + s_{i+2}$ and hence
\[
\theta_{i-1} = \tau(s_{i+1}+z_{i-1})
\le \tau(s_{i+2} + z_i)
= \theta_i.
\]

We now verify the third property of a $\kappa$-bounded encoding. Fix $i \le L-3$ with $z_i > 0$. If $f_i = 1$, the property holds by definition. Otherwise, assume $f_i = 0$. If $z_i = z^{\star}_i$, then by the properties of the set $F$, either $f_{i-1} = 1$ and $z_i \le z^{\star}_{i-1} = z_{i-1}$, or $f_{i+1} = 1$ and $z_i \le z^{\star}_{i+1} = z_{i+1}$. In either case, the required inequality holds. On the other hand, if $z_i = z^{\star}_{i+1} + s_{i+3} - s_{i+2}$, then $i+1 \in F$, and hence $f_{i+1} = 1$, so the property holds in this case as well.

It remains to verify that the encoding represents the path $(u_0,\dots,u_{L-1})$. By construction, each $u_{i+1}$ is the $a_i$-th element of $A_{\theta_i}(u_i)$, so the first condition of Definition~\ref{def:enc-represents} is satisfied. Finally, if $f_i = 1$, then $z_i = z^{\star}_i$. By the minimality of $z^{\star}_i$ (see~(\ref{eq:def-z-star})), we have for $z_i - 1$ that
\[
|A_{\tau({s_{i+2}+z_i - 1})}(u_i)| > \kappa_{z_i - 1}\,\rho_s(i+1).
\]
Observe that $\tau({s_{i+2}+z_i - 1}) = 2\theta_i$. Hence,
$A_{\tau(s_{i+2}+z_i - 1)}(u_i) = A_{2\theta_i}(u_i)$, and the second condition in Definition~\ref{def:enc-represents} is satisfied.
\end{proof}

\paragraph{Decoding and verification procedure.}
We now describe a procedure that, given an encoding $\{(z_i,a_i,f_i)\}$ and the ranks of all vertices in the tree, attempts to reconstruct a candidate path represented by this encoding. If no such path exists, the procedure fails. We will later show that the probability that the procedure succeeds is at most
$\poly(\Delta) \cdot \exp(-\sum_{i=0}^{L-3} z_i)$. This bound is sufficient to show that the sum of the probabilities of all encodings converges to $\poly(\Delta) \cdot \exp(O(L))$ as desired.

The \textsc{Decode \& Verify} procedure starts at the root of the tree, setting $u_0$ to be the root. It then iteratively constructs the path, while simultaneously verifying that the constructed sequence is indeed a candidate path. At each iteration $i < L-1$, the procedure determines $u_{i+1}$ using the triplet $(z_i,a_i,f_i)$ from the $\kappa$-bounded path encoding. Specifically, it proceeds as follows.

\medskip

\begin{tcolorbox}
\noindent\textbf{Procedure \textsc{Decode \& Verify}}

\noindent \textbf{Input: }
The root $u_0$, parameter $\kappa$,
a $\kappa$-bounded path encoding $\{(z_i,a_i,f_i)\}_{i=0}^{L-3}$ (See Definition \ref{def:path-encoding}), a sequence $s \in \calS(L)$, and vertex ranks~$r$.
\begin{enumerate}
\item Let $u_0$ be the root of the tree.

\item For each $i = 0, 1, 2, \dots L-3$ :
\begin{enumerate}
\item\label{decode:step:1} Compute $\theta_i = \tau(s_{i+2}+z_i)$.

\item\label{decode:step:3} If $f_i = 1$:
compute
\[
A_{2\theta_i}(u_i) = \textsc{Find\_Candidate\_Set}(u_i,2\theta_i),
\]
and if $|A_{2\theta_i}(u_i)| \leq \kappa_{z_i -1}\,\rho_s(i+1)$, then
terminate the algorithm with failure
(since a ``large candidate set'' was expected but not found).

\item\label{decode:step:2} Find the candidate set
\[
A_{\theta_i}(u_i) = \textsc{Find\_Candidate\_Set}(u_i,\theta_i),
\]
and set $u_{i+1}$ to be the $a_i$-th element of $A_{\theta}(u_i)$
with respect to the fixed ordering of children. If $a_i > |A_{\theta}(u_i)|$, then terminate with failure.

\item\label{decode:step:4} Finally, verify that
\[
r(u_{i+1})\in [\tau(s_{i+1}),\tau(s_{i+1}-1)).
\]
If not, terminate with failure.

\end{enumerate}

\item Return the obtained path $u_0,\dots,u_{L-2}$.
\end{enumerate}
\end{tcolorbox}

We now describe the procedure
\textsc{Find\_Candidate\_Set}$(u,\theta)$, used in the \textsc{Decode~\&~Verify}
procedure. Conceptually, this routine inspects the neighborhood
of $u$ just enough to determine which children of $u$ are promising,
and which of those belong to the candidate set $A_\theta(u)$.
Crucially, it reveals ranks only of those vertices whose ranks
must be examined in order to decide membership in $A_\theta(u)$.


\begin{tcolorbox}
\noindent\textbf{Procedure Find\_Candidate\_Set}

\noindent\textbf{Input:} a vertex $u$  and  threshold $\theta$.

\begin{enumerate}
\item\label{item:findCand:1} For each child $v \in \Gamma(u)$ do:
\begin{enumerate}
\item\label{item:findCand:1a} Inspect the ranks of the children $w \in \Gamma(v)$. For each such $w$,
if $r(w) < \theta$, reveal the value of $r(w)$; otherwise, reveal only
that $r(w) \ge \theta$. 

\item\label{item:findCand:1b} For each child $w \in \Gamma(v)$ with $r(w) < \theta$, compute $\RG(\calT_w, w, r)$. The requires revealing the ranks
of all vertices in the subtree $\calT_w$ rooted at $w$, and then simulating
RGMIS on $\calT_w$ to determine whether $w$ belongs to the independent set.
(We emphasize that the RGMIS decision for $w$ that we compute here depends only on
the ranks within the subtree rooted at $w$, $\calT_w$, so no additional
information outside this subtree is exposed. In particular, the execution above does not answer whether $w$ belongs to the independent set output by Algorithm \ref{alg:rgmis} on $w$.)

\item\label{item:findCand:1c} If $\RG(\calT_w, w, r) = \NO$ for each inspected $w$ in the above step, then mark $v$ as \emph{promising}.
\end{enumerate}

\item\label{item:findCand:2} Let $Y_u$ be the set of vertices $v \in \Gamma(u)$
that are marked as promising.

\item\label{item:findCand:3} If there exists a child $v \in \Gamma(u)$ with
$r(v) < \theta$ and $\RG(\calT_v, v, r) = \YES$,
then set $A_\theta(u) = \varnothing$. Note that the condition being checked here is whether ``$u$ is promising at threshold $\theta$'' and per Definition \ref{def:cand-set}, $A_{\theta}(u)$ is set to $\varnothing$ if this condition does not hold.

\item\label{item:findCand:4} Otherwise, define the candidate set
\[
A_\theta(u)
=
\{\, v \in Y_u :
r(v) \in [\tau({s_{\delta(v)}}), \tau({s_{\delta(v)}-1})) \,\}.
\]
Observe that conditioned on $u$ being promising at $\theta$, the above is the definition of $A_\theta(u)$ per Definition \ref{def:cand-set}.
\item Return $A_\theta(u)$.
\end{enumerate}
\end{tcolorbox}

\def\cT{\mathcal{T}}
\def\bz{\boldsymbol{z}}
\def\findCand{\textsc{Find\_Candidate\_Set}}

\begin{lemma}\label{lem:DnV-decodes-correctly}
If a $\kappa$-bounded path encoding $\bz = \{(z_i,a_i, f_i)\}_{i=0}^{L-3}$ represents a candidate path \\ $(u_0,\dots,u_{L-1})$, then the \textsc{Decode \& Verify} procedure succeeds on this encoding and returns the path $(u_0,\dots,u_{L-2})$.
\end{lemma}
\begin{proof}
\def\bz{\boldsymbol{z}}
Consider a candidate path $P = (u_0, u_1, \dots u_{L-1})$ agreeing with sequence $s \in \calS(L)$ and $\bz$ be a $\kappa$-bounded encoding of $P$ (see Definitions \ref{def:cand-path} and \ref{def:enc-represents}).
We prove that for $i \leq L-3$, given $u_i$, executing steps \ref{decode:step:1},\ref{decode:step:2},\ref{decode:step:3},\ref{decode:step:4}
computes the vertex $u_{i+1}$ correctly without failure. It follows by induction that  \textsc{Decode \& Verify} computes the desired prefix to the path, i.e., $(u_0, u_1, \dots u_{L-2})$ correctly. Step \ref{decode:step:1} of \textsc{Decode \& Verify} is identical for every call as $\theta_i$ is set from the sequence $s$ and encoding. 
At Step \ref{decode:step:2} of \textsc{Decode \& Verify}, if the set $A_{\theta}(u_i)$ is found correctly by procedure  $\findCand$, then the vertex $u_{i+1}$ is found correctly by Item \ref{def:enc-represents:a} of Definition \ref{def:enc-represents}. We next argue that $\findCand(u_i, \theta)$ correctly returns the candidate set for each~$u_i, \theta$.
\begin{claim}
Procedure $\findCand(u_i, \theta)$ correctly outputs the candidate set at $u_i$ for threshold $\theta$.
\end{claim}
\begin{proof} 
Let $A'_{\theta}(u_i)$ denote the output of $\findCand(u_i, \theta)$.
Consider an arbitrary child $v$ of $u_i$. We show that $v \in A_{\theta}(u_i)$ if and only if $v \in A'_{\theta}(u_i)$.

First observe that if $A_{\theta}(u_i)$ is forcibly set to $\emptyset$ due ``$u_i$ not being promising at threshold $\theta$'', then there exists a child $v \in \Gamma(u_i)$ with $r(v) < \theta$ satisfying $\RG(\calT_v, v, r) = \YES$, this implies that $A'_{\theta}(u_i)$ is set to $\emptyset$ as well due to Step \ref{item:findCand:3} of $\findCand$. Henceforth, we may assume that $u_i$ is promising at threshold $\theta$.

($\Rightarrow$) If $v \in A_{\theta}(u_i)$, then (i) $v$ is a child of $u_i$ that has no neighbor $w$ with $r(w) < \theta$ that belongs to the $\MIS$ in the execution of $\RG(\calT_w, w, r)$ and (ii) $r(v) \in [\tau(s_{\delta(v)}), \tau(s_{\delta(v)} - 1))$. During Step \ref{item:findCand:1}, $\RG(\calT_w, w, r)$ is executed for all the neighbors $w$ of $v$ with $r(w) < \theta$ and by choice of $v$, $w$ does not belong to the $\MIS$ during this execution.
Consequently $v$ is marked promising and since $r(v)$ lies in the expected interval is added to the candidate set in Step \ref{item:findCand:4}. 

($\Leftarrow$) On the other hand if $v \not\in A_\theta(u_i)$, then either the rank verification step at step \ref{item:findCand:4} fails or $v$ is not a promising vertex. If $v$ is not a promising vertex, then $v$ contains at least one child, say $w$ with $r(w) < \theta$ that belongs to the $\MIS$ in the execution of $\RG(\calT_w, w, r)$. Each such vertex $w$ and the simulation of $\RG(\calT_w, w, r)$ is carried out during step \ref{item:findCand:1b} attesting that $v$ is not a promising vertex and that $v$ does not belong to the set returned by $\findCand$.
\end{proof}

At Step \ref{decode:step:3}, if $f_i = 1$ then computed set $|A_{2\theta_i}(u_i)|$ must be large enough by Item \ref{def:enc-represents:b} of Definition \ref{def:enc-represents}.
At Step \ref{decode:step:4}, the condition $r(u_{i+1}) \in[\tau(s_{i+1}), \tau(s_{i+1}-1))$ holds by the definition of candidate path. 
Thus, the call to \textsc{Decode \& Verify} when executed with an encoding of the path $P$ correctly computes $u_{i+1}$ given that $u_i$ is found correctly, for every $i \leq L-3$.
\end{proof}

We now proceed to analyze the probability that \textsc{Decode \& Verify} succeeds on a given $\kappa$-bounded path encoding. The high-level idea underlying the analysis is as follows. Whenever the flag $f_i$ is set (i.e., $f_i = 1$), the procedure verifies that a low-probability event, namely $\{|A_{2\theta}(u_i)| > \kappa_{z_i-1}\,\rho_s(i+1)\}$, has occurred. We use Lemma~\ref{lem:prob-ub-large-candidate-set} to show that the probability of this event (conditioned on the first $i-1$ steps of \textsc{Decode \& Verify}) is approximately $\exp(-\Theta(z_i))$. Consequently, the probability that the procedure does not fail at step $i$ is at most $\exp(-\Theta(z_i))$, and the probability that \textsc{Decode \& Verify} succeeds at every step is at most $\exp(-\Theta(\sum_{i : f_i = 1} z_i))$. We formalize this intuition in Lemma~\ref{lem:prob-path-encoding-valid}, but first carefully examine the randomness revealed during  the first $i$ steps of \textsc{Decode \& Verify}.

\def\bz{\boldsymbol{z}}
\begin{lemma}\label{lem:DV-revealed-vertices}
Let $i \in \{0, 1,\dots,L-2\}$ with $f_i = 1$. During steps
$0,\dots,i-1$ of the execution of the \textsc{Decode \& Verify} procedure on a valid $\kappa$-bounded encoding,
for every child
$v$ of $u_i$ with $r(v) \ge \theta_i$, no information about the ranks of
vertices in the sub-tree $\calT_v - v$ is revealed. For the rank $r(v)$, it is only revealed completely if $r(v) < \theta'$ and otherwise, conditioned on the execution of Decode\&Verify so far, $r(v)$ is uniformly distributed in $[\theta', 1)$ for some $\theta' \leq \theta$.
\end{lemma}

\begin{proof}
We examine which ranks may be revealed during iterations $0,\dots,i-1$ of the
\textsc{Decode \& Verify} procedure. The only operations that may reveal
the ranks of vertices other than the path vertices $u_0,\dots,u_i$ are calls
to \textsc{Find\_Candidate\_Set}. During iteration $j$, the procedure invokes
\textsc{Find\_Candidate\_Set} once with arguments $(u_j,\theta_j)$ if $f_j=0$,
and twice with arguments $(u_j,\theta_j)$ and $(u_j,2\theta_j)$ if $f_j=1$.

First, fix a step $j \le i-1$ and consider a call to
\textsc{Find\_Candidate\_Set}$(u_j,\theta_j)$. By definition of this procedure, for each child $v \in \Gamma(u_j)$ it inspects
children $w' \in \Gamma(v)$ and completely reveals $r(w')$
only if $r(w') < \theta_j$, where $\theta_j$ is the threshold used at step $j$. If this condition holds, the procedure simulates the RGMIS rule inside the subtree $\calT_{w'}$ rooted at $w'$ and reveals all the ranks of vertices in this subtree. Observe that the rank of vertex $u_{j+2}$ satisfies
$$
r(u_{j+2}) \geq \tau({s_{j+2}})
\geq \tau({s_{j+2} + z_j}) = \theta_j.
$$
Moreover, if the flag $f_j=1$, then 
$z_j \geq 1$ and
$$
r(u_{j+2}) \geq \tau(s_{j+2})
\geq \tau(s_{j+2} +z_j-1) = 2\theta_j.
$$

Thus, in both cases, the calls to $\findCand$ at $u_j$ do not reveal ranks of vertices beneath any grandchild $w$ of $u_j$ with $r(w) \geq \theta_j$ when $f_j \neq 1$ and $r(w) \geq 2 \theta_j$ when $f_j = 1$. In both cases $u_{j+2}$ is included as desired.

Thus, when $j < i-1$, the rank of every vertex in $\cT_{u_{j+2}}$ except $u_{j+2}$ has not been queried.
If the calls to $\findCand$ at $u_j$ inspected the rank of some grandchild $w$ of $u_j$ with $r(w) \geq \theta_j(1 + f_j)$, then it only inspects whether $r(w) \geq (1 + f_j) \theta_j$. 

We next consider the case $j = i-1$. Note that $i \geq 1$. Since $f_i = 1$, $f_{i-1} = 0$ and therefore the calls to $\findCand$ at $u_{i-1}$ only tests whether $r(v) \geq \theta_{i-1} (1 + f_{i-1}) = \theta_{i-1}$. If the condition passes, then no vertex beneath $\cT_v$ is queried.
The requirement $f_i = 1$ enforces that $\theta_{i-1} \leq \theta_i$ by Item \ref{cond:item:c} of Definition \ref{def:path-encoding} and thus $r(v) \geq \theta_i \geq \theta_{i-1}$, so the test passes and no vertex beneath $\cT_v$ is queried. 
This completes the proof.

\end{proof}

\begin{lemma}\label{lem:prob-path-encoding-valid}
The probability that a given $\kappa$-bounded path encoding
$\{(z_i,a_i,f_i)\}$ represents a candidate path is at most
\[
\Pr\bigl(\textsc{Decode \& Verify} \;\;\text{\emph{succeeds}}\bigr)
\le 
\exp\Bigl(-3\,\sum_{i:f_i=1}z_i\Bigr).
\]
\end{lemma}

\begin{proof}
We run the \textsc{Decode \& Verify} procedure step by step and, for each
$i$, bound the probability that step $i$ succeeds conditioned on all previous
steps having succeeded. We then multiply these bounds. If $f_i = 0$, we use the
trivial upper bound~$1$ for the conditional success probability at step~$i$.

Now fix a step $i$ with $f_i = 1$. Before bounding the probability that step $i$
succeeds, we first argue that Lemma~\ref{lem:prob-ub-large-candidate-set}
applies to the random set $A_{2\theta_i}(u_i)$.
By Lemma~\ref{lem:DV-revealed-vertices}, during steps $0,\dots,i-2$ the
procedure does not reveal any ranks in the subtree $\calT_{u_i}$ except possibly
$r(u_i)$ itself. Moreover, at step $i-1$, it may only 
\begin{itemize}
\item reveal the ranks of vertices in subtrees $\calT_v$ for
$v \in \Gamma(u_i)$ with $r(v) < \theta_i$;
\item partially reveal the ranks of vertices $v \in \Gamma(u_i)$ with
$r(v) \ge \theta_i$, i.e., reveal only that $r(v) \ge \theta_i$. It does not
reveal any ranks of vertices in $\calT_v$ other than possibly $v$ itself.
\end{itemize}

Thus, the revealed randomness immediately before step $i$ satisfies the
conditions of Lemma~\ref{lem:prob-ub-large-candidate-set}. Moreover, since $f_i = 1$, by Definition~\ref{def:path-encoding} we have $z_i \ge 1$, and hence
\[
\theta_i = \tau({s_{i+2}+z_i})\le \tau({s_{i+2}+1}) = \tau({s_{i+2}})/2.
\]
Therefore, the assumptions of Lemma~\ref{lem:prob-ub-large-candidate-set}
are satisfied, and by this lemma, for every  $T>0$, we have
\[
\Pr\bigl(|A_{2\theta_i}(u_i)| \ge T\bigr)
\le \exp\!\left(-\,\frac{T}{\frac{\tau({s_{\delta(u_i)+1}})}{\theta_i} + 1}\right).
\]
We apply this bound with
$T=\lceil \kappa_{z_i-1}\rho_s(i+1)\rceil$ to upper bound the probability that
the \textsc{Decode \& Verify} procedure does not fail at step $i$:
\begin{align*}
\Pr\bigl(|A_{2\theta_i}(u_i)| \ge \kappa_{z_i-1}\rho_s(i+1)\bigr)
&\le \exp\!\left(-\,\frac{\kappa_{z_i-1}\rho_s(i+1)}{\frac{\tau(s_{i+1})}{\tau(s_{i+2}+z_i)} + 1}\right).
\end{align*}
Then,
\[
\frac{\tau(s_{i+1})}{\tau(s_{i+2}+z_i)}
= 2^{s_{i+2}-s_{i+1}} \cdot 2^{z_i}
= \rho_s(i+1)\cdot 2^{z_i}.
\]
Substituting into the bound, we obtain
\[
\Pr\bigl(|A_{2\theta_i}(u_i)| \ge \kappa_{z_i-1}\rho_s(i+1)\bigr)
\le \exp\!\left(
-\,\frac{\kappa_{z_i-1}\rho_s(i+1)}{\rho_s(i+1)\cdot 2^{z_i} + 1}
\right).
\]
Since $\rho_s(i+1) \ge 1$ and $z_i \geq 1$, we have
\[
\rho_s(i+1)\cdot 2^{z_i} + 1 \le \nicefrac{3}{2}\,\rho_s(i+1)\cdot 2^{z_i}.
\]
Since $\kappa_{z_i-1} = 5\cdot z_i\cdot 2^{z_i}$, we get
\[
\frac{\kappa_{z_i-1}\,\rho_s(i+1)}{\rho_s(i+1)\cdot 2^{z_i} + 1}
\ge \frac{5\cdot z_i\cdot 2^{z_i}\cdot \rho_s(i+1)}{\nicefrac{3}{2}\,\rho_s(i+1)\cdot 2^{z_i}}
= \frac{10z_i}{3}
\]
It follows that
\[
\Pr\bigl(|A_{2\theta_i}(u_i)| \ge \kappa_{z_i-1}\rho_s(i+1)\bigr)
\le e^{-\nicefrac{10}{3}\,z_i}.
\]

Thus, the conditional probability that step $i$ succeeds, given that all
previous steps have succeeded, is at most
$e^{-\nicefrac{10}{3}\,z_i}$ whenever $f_i=1$, while for $f_i=0$
we use the trivial bound $1$. Multiplying these bounds yields
\[
\Pr(\textsc{Decode \& Verify}\;\text{\emph{succeeds}})
\le \prod_{i:f_i=1}
e^{-\nicefrac{10}{3}\,z_i}\leq 
\prod_{i:f_i=1}
e^{-3\,z_i},
\]
as claimed.
\end{proof}
Lemma~\ref{lem:prob-path-encoding-valid} establishes a strong upper bound on the probability that a bounded encoding corresponds to a candidate path, which we will use to estimate the expected number of candidate paths of length $L$. However, the bound depends on the sum $\sum_{i : f_i = 1} z_i$, which is not convenient to work with directly. We thus show how to replace this term with the more tractable expression $ \tfrac{1}{3} \left(\sum_{i} z_i \right)
- \tfrac{m}{3}$.

\begin{lemma}
\label{lem:sum-of-zs}
For every valid $\kappa$-bounded encoding, it holds that
\[
\sum\limits_{i=0}^{L-3} z_i \leq m + 3 \cdot \sum\limits_{i : f_i = 1} z_i.
\]
\end{lemma}

\begin{proof}
We upper bound the sum $\sum\limits_{i=0}^{L-3} z_i$ by bounding each term individually. We ignore indices $i$ with $z_i = 0$, since they do not affect the sum, and upper bound each $z_i$ with $f_i = 1$ by itself. We now consider indices with $z_i > 0$ and $f_i = 0$. 
By Item~\ref{cond:item:d} of Definition~\ref{def:path-encoding}, each such $z_i$ is upper bounded by either $z_{i-1}$ or $z_{i+1} + s_{i+3} - s_{i+2}$. Moreover, in the former case $f_{i-1} = 1$, and in the latter case $f_{i+1} = 1$. Replacing each such $z_i$ by its corresponding upper bound, we obtain an upper bound on the total sum.

The resulting expression consists of terms of the form $z_i$ and $z_{i} + s_{i+2} - s_{i+1}$ for indices with $f_i = 1$. Observe that each $z_i$ appears with coefficient at most $3$, depending on whether it is used to upper bound only itself, itself and one of $z_{i-1}$ or $z_{i+1}$, or all three of $z_i$, $z_{i-1}$, and $z_{i+1}$. For each $i$ with $f_i = 1$, the term $s_{i+2} - s_{i+1}$ may appear with coefficient $1$, or not appear at all, depending on whether $z_i$ is used to upper bound $z_{i-1}$. Thus,
\[
\sum\limits_{i=0}^{L-3} z_i \leq \sum_{i : f_i = 1} 3 z_i + \sum_{i : f_i = 1} (s_{i+2} - s_{i+1}).
\]

Since $(s_i)$ is a non-decreasing sequence, we have $s_{i+2} - s_{i+1} \geq 0$ for all $i$, and hence the second sum is upper bounded by
\[
\sum\limits_{i=0}^{L-3} (s_{i+2} - s_{i+1}) = s_{L-1} - s_1 \leq m.
\]
This completes the proof.
\end{proof}

\begin{corollary}\label{cor:lem:prob-path-encoding-valid}
The probability that a $\kappa$-bounded path encoding $\{(z_i,a_i,f_i)\}$ represents at least one candidate path is at most
\[
e \cdot \Delta^{\log_2 e} \cdot 
\exp\Bigl(-\sum_{i=0}^{L-3} z_i\Bigr).
\]
Moreover, it represents at most $\Delta$ candidate paths.
\end{corollary}
\begin{proof}
If a $\kappa$-bounded path encoding represents a candidate path, then the \textsc{Decode \& Verify} procedure must be able to successfully decode and verify the encoding (see~Lemma~\ref{lem:DnV-decodes-correctly}). By 
Lemmas~\ref{lem:prob-path-encoding-valid} and 
\ref{lem:sum-of-zs}, we have that for every $\kappa$-bounded encoding $\{(z_i,a_i,f_i)\}$:
\begin{align*}
\Pr\bigl(\textsc{Decode \& Verify}\,\text{succeeds}\bigr) 
&\leq
\exp\Bigl(-3\sum\limits_{i : f_i = 1} z_i
\Bigr)\\
&\leq 
\exp\Bigl(
-3\bigl(-\tfrac{m}{3} + \tfrac{1}{3}\sum\limits_{i=0}^{L-3} z_i\bigr)
\Bigr)
\\ 
&= \exp(m) \cdot \exp\Bigl(
-\sum\limits_{i=0}^{L-3} z_i
\Bigr).
\end{align*}
Substituting $m \leq \log_2 \Delta + 1$ completes the proof of the probability bound.

\def\bz{\boldsymbol{z}}
From Lemma \ref{lem:DnV-decodes-correctly}, the \textsc{Decode \& Verify} procedure executed with $\bz = \{(z_i, a_i, f_i)\}$  constructs a prefix $(u_0, u_1, \dots u_{L-2})$ such that every candidate path encoded by $\bz$ has the same prefix. There are at most $\Delta$ choices for $u_{L-1}$ as a neighbor of $u_{L-2}$, and thus the encoding can represent at most $\Delta$ such paths.
\end{proof}

We now prove Lemma~\ref{lem:bound-candidate-paths} with a specific choice of constants $C_1$ and $C_2$.

\begin{lemma}
For every $s \in \calS(L)$, the expected number of candidate paths of length $L$ agreeing with $s$ is at most
\[
f(L,\Delta) = (2e)\cdot \Delta^{2 + \log_2 e} \cdot 290^{\max(0, L-2)}.
\]
\end{lemma}
\begin{proof}
Note that when $L \leq 2$, the bound in the lemma is trivial, as there are at most $\Delta^{L-1}$ paths of $L$ vertices starting from $u_0$. Henceforth, we assume $L \geq 3$.

We first give a high-level overview of the proof. We first bound the expected number of encodings $\bz = \{(z_i, a_i, f_i)\}$ that pass the \textsc{Decode \& Verify} procedure. Each encoding can refer to multiple candidate paths. If $P$ is a candidate path for some encoding $\bz$, then from Lemma \ref{lem:DnV-decodes-correctly}, all the vertices of $P$ except the last vertex can be determined by executing the \textsc{Decode \& Verify} procedure which is deterministic for a fixed rank $r$. Hence the candidate paths encoded by $\bz$ can only differ in the last vertex of the path. This implies that there are at most $\Delta$ paths encoded by $\bz$. The expected number of encodings is bound in a straightforward way, i.e., sum the probability bound by Corollary \ref{cor:lem:prob-path-encoding-valid} over all possible encodings.
The number of encodings for a fixed set $(z_0, z_1, \dots z_{L-3})$ is, $\poly(\Delta) \prod\limits_{i=0}^{L-3} O(z_i 2^{z_i})$. The probability of any such encoding passing the \textsc{Decode \& Verify} test is $\exp(-\sum z_i)$, which is exponentially smaller than $\prod z_i 2^{z_i}$. Thus, the weighted sum of all encodings where the weights are the probabilities converges to $\poly(\Delta)\cdot \exp(O(L))$.

Fix the sequence $(z_0,\dots,z_{L-3})$. For every $i$, once $z_i$ is fixed, the number
of possible values of $a_i$ is at most
\[
\kappa_{z_i}\rho_s(i+1).
\]
Hence, for this fixed sequence $(z_i)$, the total number of possible choices of the
indices $(a_i)$ is at most
\[
\prod_{i=0}^{L-3} \kappa_{z_i}\rho_s(i+1)
=
\prod_{i=0}^{L-3} \rho_s(i+1)
\cdot
\prod_{i=0}^{L-3} \kappa_{z_i}.
\]
Using the definition of $\rho_s(i)$, we obtain
\[
\prod_{i=0}^{L-3} \kappa_{z_i}\rho_s(i+1)
=
\frac{\tau({s_1})}{\tau({s_{L-1}})}
\cdot \prod_{i=0}^{L-3} \kappa_{z_i}
\leq
2\Delta \cdot \prod_{i=0}^{L-3} \kappa_{z_i}.
\]

The number of possible flag sequences $(f_i)$ is at most $2^{L-2}$. Therefore, the
total number of $\kappa$-bounded path encodings with a prescribed sequence
$(z_0,\dots,z_{L-3})$ is at most
\[
\Delta \cdot 2^{L-1}\cdot \prod_{i=0}^{L-3} \kappa_{z_i}.
\]
By Corollary~\ref{cor:lem:prob-path-encoding-valid}, the probability that such an encoding
represents a candidate path is at most 
\[
e\Delta^{\log_2 e} \exp\Bigl(-\sum_{i=0}^{L-3} z_i\Bigr).
\]
It follows that the expected number of encodings with prescribed values
$(z_0,\dots,z_{L-3})$ that represent a candidate path is at most
\[
e\cdot \Delta^{1 + \log_2 e} \cdot 2^{L-1}\cdot 
\prod_{i=0}^{L-3} \bigl(\kappa_{z_i} e^{-z_i}\bigr).
\]
Summing over all possible choices of $(z_i)$, we obtain
\begin{align*}
e\cdot \Delta^{1 + \log_2 e} \cdot 2^{L-1}\cdot 
\sum_{z_0,\dots,z_{L-3}\in \bbZ_{\geq 0}}
\prod_{i=0}^{L-3} \kappa_{z_i} e^{-z_i}
=
e\cdot \Delta^{1 + \log_2 e} \cdot 2^{L-1}\cdot 
\left(
\sum_{z\in \bbZ_{\geq 0}}
\kappa_{z} e^{-z}
\right)^{L-2}.
\end{align*}
The sum on the right-hand side is upper bounded as follows:
\[
\sum_{z\in \bbZ_{\geq 0}}
\kappa_{z} e^{-z} 
=
\sum_{z\in \bbZ_{\geq 0}}
5(z+1)2^{z+1} e^{-z}
=
10\sum_{z\in \bbZ_{\geq 0}}
(z+1)\left(\frac{2}{e}\right)^{z}.
\]
Using the  Taylor series  of $\frac{1}{(1-x)^2}$ and substituting $x=\frac{2}{e}$, we obtain
\[
\sum_{z\in \bbZ_{\geq 0}}
\kappa_{z} e^{-z}
=
\frac{10e^2}{(e-2)^2}\leq 145.
\]
Each encoding can correspond to at most $\Delta$ candidate paths, since $u_{L-1}$ is the only vertex that is not identified by the Decode\&Verify procedure and there are $\Delta$ choices for $u_{L-1}$ as a child of $u_{L-2}$.
Therefore, the expected number of encodings that represent a candidate path is at most
\[
e\cdot \Delta^{2 + \log_2 e} \cdot 2^{L-1}\cdot 145^{L-2}
=
2e\cdot \Delta^{2 + \log_2 e} \cdot 290^{L-2}.
\]
as desired.
\end{proof}

Finally each $\kappa$-bounded path encoding represents at most $\Delta$ candidate paths. 
The desired constants $C_1, C_2$ for Lemma \ref{lem:bound-candidate-paths} are bounded by $C_1 \leq 290$ and $C_2 \le 3 + (1/\ln 2) < 4.5$. Hence, we obtain a bound of $O(\Delta^{425})$ on the expected size of the query tree generated by Algorithm \ref{alg:rgmis}.
\bibliographystyle{alpha}
\bibliography{references}

\appendix

\section{Deferred proofs}
\subsection{Bounding low-rank paths}
\label{sec:low-rank-paths}
Consider a tree $\calT$, a vertex $u \in T$ and the execution of $\RG(\calT, u, r)$ over uniformly random $r$. We show that the expected number of query paths to vertices with rank $[\tau(m), 1)$ is at least a constant factor of the total expected number of query paths.

    \def\qlow{Q_{\text{low}}}
\begin{lemma}[Bounding expected low-rank paths]
\label{lem:bound-low-rank-paths}
    For an arbitrary vertex $v$ in an arbitrary tree $\calT$ of degree at most $\Delta$, let $\qlow(T, v)$ denote the set of query paths $P$ in the execution of $\RG(\calT, v)$ such that every vertex $w \in P - v$ has $r(w) < 1/\Delta$. It holds that,
    \[
        \mathbbm{E}_r{[|\qlow(T, v)| \  \mid r(v)]} \leq e
    \]
\end{lemma}
\begin{proof}
    The number of paths with $L$ edges is at most $\Delta^L$. For each such path, the probability that the ranks of all the vertices in it except for $v$ lie in $[0, \tau(m))$ is at most $\Delta^{-L}$. Given that all ranks lie in $[0, \tau(m))$, the probability that the path is $r$-monotone is $1/L!$.
    Thus, the expected number of such paths is at most $\sum\limits_{L \geq 0} \Delta^L \cdot \Delta^{-L} \cdot 1/L! = e$.
\end{proof}

\begin{lemma}[Reduction to high-rank paths]
\label{lem:deal-with-low-rank}
    Consider an arbitrary tree $\calT$ and vertex $u \in \calT$.
    Let $X$ be the expected number of query paths in the execution of $\RG(\calT, u)$ that agree with some sequence $s \in \calS$. In particular, this implies that each vertex on the path has rank at least $\tau(m)$.
    It holds that the total expected number of query paths in $\RG(\calT, u)$ is at most $e(X + 1)$.
\end{lemma}
\def\path{\text{path}}
\begin{proof}
    For convenience, we add an imaginary vertex $\perp$ to the tree $\calT_\perp$ whose only neighbor is $u$ and root the new tree $\calT_\perp$ at $\perp$. We consider $r(\perp) = 1 \geq \tau(m)$. We make this change so that every query path in $\RG(\calT_{\perp}, \perp, r)$ begins at a vertex with rank at least $\tau(m)$ with probability $1$. At the same time, the size of the query tree produced by $\RG(\calT_\perp, \perp, r)$ is same as the size of the desired query tree $\RG(\calT, u, r)$ after ignoring the vertex $\perp$.
    
    To prove the lemma, consider an arbitrary vertex $v \in \calT_\perp$.
    Let $\path(v)$ denote the unique path from $\perp$ to $v$ in $\calT_\perp$.
    Let $E_v$ denote the event that $\RG(\calT_\perp, u)$ queries $v$ and $r(v) \geq \tau(m)$. Let $p(v) = \Pr[E_v]$. 
    \begin{align*}
        E_v &: \RG(\calT_\perp, \perp, r) \text{ queries } v \land r(v) \geq \tau(m) \\
        p(v) &= \Pr[E_v]
    \end{align*}
    Note that $E_{\perp}$ is a sure event, i.e., $p(\perp) = 1$.
    Note that $E_v$ is independent of the ranks of vertices in the sub-tree $\calT_v$ excluding $v$.
    To see this, observe that in the procedure $\RG(\calT_\perp, \perp, r)$ when $v$ is visited for the first time, the entire sub-tree beneath $v$ has been unexplored, and hence their ranks have not been read.

    Observe that by linearity of expectation, 
    \begin{align}
    \label{eqn:expression-for-x}
    X = \sum\limits_{v \in \calT} p(v).
    \end{align}

    Let $\qlow(v)$ denote the set of query paths in $\RG(\calT_v, v)$ such that all of the vertices in the path have rank in $[0, \tau(m))$ except $v$. 
    We have, $\mathbbm{E}[|{\qlow}(v)| \mid r(v)] \leq e$ by Lemma \ref{lem:bound-low-rank-paths} and the independence of ranks in $\calT_v$.
    Finally, the expected total number of paths $Q(\calT_\perp, \perp)$ is given by,
    \begin{align*}
        \mathbbm{E}_r[|Q(\calT_\perp, \perp)|] &= \sum\limits_{v \in \calT_\perp} \Pr[E_v] \cdot \mathbbm{E}_r[|\qlow(\calT_v, v)| \mid E_v] \ \ \ \ \text{[Using linearity of expectation} \text{]}  \\
        &= \sum\limits_{v \in \calT_\perp} \Pr[E_v] \cdot \mathbbm{E}_r[|\qlow(\calT_v, v)| \mid r(v)] \ \ \ \ \text{[Using independence of } E_v \text{ from } \calT_v - v \text{]} \\
        &\leq e \sum\limits_{v \in \calT_\perp} \Pr[E_v] \ \ \ \ \text{[Using Lemma } \ref{lem:bound-low-rank-paths} \text{]} \\
        &= e (X + 1) \ \ \ \ \text{[By Equation } \ref{eqn:expression-for-x} \text{, with +1 accounting for} \perp \text{]}
    \end{align*}
    as desired.
\end{proof}

%% file: references.bib
@inproceedings{nguyen2008constant,
  title={Constant-time approximation algorithms via local improvements},
  author={Nguyen, Huy N and Onak, Krzysztof},
  booktitle={Proceedings of the 49th Annual IEEE Symposium on Foundations of Computer Science (FOCS 2008)},
  pages={327--336},
  year={2008},
  organization={IEEE}
}

@inproceedings{kapralov2020space,
  title={Space-efficient approximation to maximum matching size from uniform edge samples},
  author={Kapralov, Michael and Mitrovi{\'c}, Slobodan and Norouzi-Fard, Ashkan and Tardos, Jakab},
  booktitle={Proceedings of the 14th Annual ACM-SIAM Symposium on Discrete Algorithms (SODA 2020)},
  pages={1753--1772},
  year={2020},
  organization={SIAM}
}

@inproceedings{yoshida2009improved,
  title={An improved constant-time approximation algorithm for maximum matchings},
  author={Yoshida, Yuichi and Yamamoto, Masaki and Ito, Hiro},
  booktitle={Proceedings of the 41st Annual ACM Symposium on Theory of Computing (STOC 2009)},
  pages={225--234},
  year={2009}
}

@inproceedings{dalirrooyfard2026simple,
  title={A simple average-case analysis of recursive randomized greedy {MIS}},
  author={Dalirrooyfard, Mina and Makarychev, Konstantin and Mitrovi{\'c}, Slobodan},
  booktitle={Proceedings of the SIAM Symposium on Simplicity in Algorithms (SOSA 2026)},
  pages={241--246},
  year={2026},
  organization={SIAM}
}

@inproceedings{onak2012near,
  title={A near-optimal sublinear-time algorithm for approximating the minimum vertex cover size},
  author={Onak, Krzysztof and Ron, Dana and Rosen, Michal and Rubinfeld, Ronitt},
  booktitle={Proceedings of the 23rd Annual ACM-SIAM Symposium on Discrete Algorithms (SODA 2012)},
  pages={1123--1131},
  year={2012},
  organization={SIAM}
}

@inproceedings{behnezhad2022time,
  title={Time-optimal sublinear algorithms for matching and vertex cover},
  author={Behnezhad, Soheil},
  booktitle={Proceedings of the 62nd Annual IEEE Symposium on Foundations of Computer Science (FOCS 2021)},
  pages={873--884},
  year={2021},
  organization={IEEE}
}

@inproceedings{bhattacharya2023dynamic,
  title={Dynamic $(1+\epsilon)$-approximate matching size in truly sublinear update time},
  author={Bhattacharya, Sayan and Kiss, Peter and Saranurak, Thatchaphol},
  booktitle={Proceedings of the 64th Annual IEEE Symposium on Foundations of Computer Science (FOCS 2023)},
  pages={1563--1588},
  year={2023},
  organization={IEEE}
}

@inproceedings{behnezhad2023dynamic,
  title={Dynamic algorithms for maximum matching size},
  author={Behnezhad, Soheil},
  booktitle={Proceedings of the Annual ACM-SIAM Symposium on Discrete Algorithms (SODA 2023)},
  pages={129--162},
  year={2023},
  organization={SIAM}
}

@inproceedings{mahabadi2025sublinear,
  title={Sublinear metric Steiner tree via improved bounds for set cover},
  author={Mahabadi, Sepideh and Roghani, Mohammad and Tarnawski, Jakub and Vakilian, Ali},
  booktitle={Proceedings of the Innovations in Theoretical Computer Science Conference (ITCS 2025)},
  pages={74--1},
  year={2025},
  organization={Schloss Dagstuhl--Leibniz-Zentrum f{\"u}r Informatik}
}

@inproceedings{mahabadi2026sublinear,
  title={Sublinear metric Steiner forest via maximal independent set},
  author={Mahabadi, Sepideh and Roghani, Mohammad and Tarnawski, Jakub and Vakilian, Ali},
  booktitle={Proceedings of the Annual ACM-SIAM Symposium on Discrete Algorithms (SODA 2026)},
  pages={5941--5959},
  year={2026},
  organization={SIAM}
}

@inproceedings{mahabadi20250,
  title={A 0.51-approximation of maximum matching in sublinear $n^{1.5}$ time},
  author={Mahabadi, Sepideh and Roghani, Mohammad and Tarnawski, Jakub},
  booktitle={Proceedings of the International Colloquium on Automata, Languages, and Programming (ICALP 2025)},
  pages={116--1},
  year={2025},
  organization={Schloss Dagstuhl--Leibniz-Zentrum f{\"u}r Informatik}
}

@inproceedings{das2026history,
  title={History-independent maximal matchings can be surprisingly efficient, and lead to better worst-case guarantees},
  author={Das, Rathish and Kuszmaul, William},
  booktitle={Proceedings of the Annual ACM-SIAM Symposium on Discrete Algorithms (SODA 2026)},
  pages={3999--4022},
  year={2026},
  organization={SIAM}
}

@article{varma2023average,
  title={Average sensitivity of graph algorithms},
  author={Varma, Nithin and Yoshida, Yuichi},
  journal={SIAM Journal on Computing},
  volume={52},
  number={4},
  pages={1039--1081},
  year={2023},
  publisher={SIAM}
}

@inproceedings{azarmehr2025stochastic,
  title={Stochastic matching via in-n-out local computation algorithms},
  author={Azarmehr, Amir and Behnezhad, Soheil and Ghafari, Alma and Rubinfeld, Ronitt},
  booktitle={Proceedings of the Annual ACM Symposium on Theory of Computing (STOC 2025)},
  pages={1055--1066},
  year={2025}
}

@inproceedings{behnezhad2022almost,
  title={Almost 3-approximate correlation clustering in constant rounds},
  author={Behnezhad, Soheil and Charikar, Moses and Ma, Weiyun and Tan, Li-Yang},
  booktitle={Proceedings of the 63rd Annual IEEE Symposium on Foundations of Computer Science (FOCS 2022)},
  pages={720--731},
  year={2022},
  organization={IEEE}
}

@inproceedings{ghaffari2022local,
  title={Local computation of maximal independent set},
  author={Ghaffari, Mohsen},
  booktitle={2022 IEEE 63rd Annual Symposium on Foundations of Computer Science (FOCS)},
  pages={438--449},
  year={2022},
  organization={IEEE}
}

@article{rubinfeld2011fast,
  title={Fast local computation algorithms},
  author={Rubinfeld, Ronitt and Tamir, Gil and Vardi, Shai and Xie, Ning},
  journal={arXiv preprint arXiv:1104.1377},
  year={2011}
}

@inproceedings{alon2012space,
  title={Space-efficient local computation algorithms},
  author={Alon, Noga and Rubinfeld, Ronitt and Vardi, Shai and Xie, Ning},
  booktitle={Proceedings of the twenty-third annual ACM-SIAM symposium on Discrete Algorithms},
  pages={1132--1139},
  year={2012},
  organization={SIAM}
}

@article{even2018best,
  title={Best of two local models: Centralized local and distributed local algorithms},
  author={Even, Guy and Medina, Moti and Ron, Dana},
  journal={Information and Computation},
  volume={262},
  pages={69--89},
  year={2018},
  publisher={Elsevier}
}

@inproceedings{levi2015local,
  title={Local computation algorithms for graphs of non-constant degrees},
  author={Levi, Reut and Rubinfeld, Ronitt and Yodpinyanee, Anak},
  booktitle={Proceedings of the 27th ACM symposium on Parallelism in Algorithms and Architectures},
  pages={59--61},
  year={2015}
}

@article{levi2017centralized,
  title={A (centralized) local guide},
  author={Levi, Reut and Medina, Moti and others},
  journal={Bulletin of EATCS},
  volume={2},
  number={122},
  year={2017}
}

@article{gamarnik2010randomized,
  title={Randomized greedy algorithms for independent sets and matchings in regular graphs: Exact results and finite girth corrections},
  author={Gamarnik, David and Goldberg, David A},
  journal={Combinatorics, Probability and Computing},
  volume={19},
  number={1},
  pages={61--85},
  year={2010},
  publisher={Cambridge University Press}
}

@article{brightwell2017greedy,
  title={The greedy independent set in a random graph with given degrees},
  author={Brightwell, Graham and Janson, Svante and Luczak, Malwina},
  journal={Random Structures \& Algorithms},
  volume={51},
  number={4},
  pages={565--586},
  year={2017},
  publisher={Wiley Online Library}
}

@article{krivelevich2024greedy,
  title={Greedy maximal independent sets via local limits},
  author={Krivelevich, Michael and M{\'e}sz{\'a}ros, Tam{\'a}s and Michaeli, Peleg and Shikhelman, Clara},
  journal={Random Structures \& Algorithms},
  volume={64},
  number={4},
  pages={986--1015},
  year={2024},
  publisher={Wiley Online Library}
}

@article{contat2022surprising,
  title={Surprising identities for the greedy independent set on Cayley trees},
  author={Contat, Alice},
  journal={Journal of Applied Probability},
  volume={59},
  number={4},
  pages={1042--1058},
  year={2022},
  publisher={Cambridge University Press}
}

@inproceedings{karp1972reducibility,
  author       = {Richard M. Karp},
  editor       = {Raymond E. Miller and
                  James W. Thatcher},
  title        = {Reducibility Among Combinatorial Problems},
  booktitle    = {Proceedings of a symposium on the Complexity of Computer Computations,
                  held March 20-22, 1972, at the {IBM} Thomas J. Watson Research Center,
                  Yorktown Heights, New York, {USA}},
  series       = {The {IBM} Research Symposia Series},
  pages        = {85--103},
  publisher    = {Plenum Press, New York},
  year         = {1972},
  url          = {https://doi.org/10.1007/978-1-4684-2001-2\_9},
  doi          = {10.1007/978-1-4684-2001-2\_9},
  bibsource    = {dblp computer science bibliography, https://dblp.org}
}

@article{flory1939intramolecular,
  title={Intramolecular reaction between neighboring substituents of vinyl polymers},
  author={Flory, Paul J},
  journal={Journal of the American Chemical Society},
  volume={61},
  number={6},
  pages={1518--1521},
  year={1939},
  publisher={ACS Publications}
}

@article{evans1993random,
  title={Random and cooperative sequential adsorption},
  author={Evans, James W},
  journal={Reviews of modern physics},
  volume={65},
  number={4},
  pages={1281},
  year={1993},
  publisher={APS}
}
